\documentclass[preprint2,times,tighten]{aastex6}
\usepackage{mathptmx}   

\usepackage{etoolbox}
\makeatletter
\patchcmd{\NAT@citex}
  {\@citea\NAT@hyper@{%
     \NAT@nmfmt{\NAT@nm}%
     \hyper@natlinkbreak{\NAT@aysep\NAT@spacechar}{\@citeb\@extra@b@citeb}%
     \NAT@date}}
  {\@citea\NAT@nmfmt{\NAT@nm}%
   \NAT@aysep\NAT@spacechar\NAT@hyper@{\NAT@date}}{}{}
\patchcmd{\NAT@citex}
  {\@citea\NAT@hyper@{%
     \NAT@nmfmt{\NAT@nm}%
     \hyper@natlinkbreak{\NAT@spacechar\NAT@@open\if*#1*\else#1\NAT@spacechar\fi}%
       {\@citeb\@extra@b@citeb}%
     \NAT@date}}
  {\@citea\NAT@nmfmt{\NAT@nm}%
   \NAT@spacechar\NAT@@open\if*#1*\else#1\NAT@spacechar\fi\NAT@hyper@{\NAT@date}}
  {}{}
\makeatother

\begin{document}

\title{The evolution of gas giant entropy during formation by runaway accretion}
\author{David Berardo\altaffilmark{1,2}, Andrew Cumming\altaffilmark{1,2}, Gabriel-Dominique Marleau\altaffilmark{3}}
\affil{
\altaffilmark{1}{Department of Physics and McGill Space Institute, McGill University, 3550 rue University, Montreal, QC, H3A 2T8, Canada}\\
\altaffilmark{2}{Institut de recherche sur les exoplan\`{e}tes (iREx)}\\
\altaffilmark{3}{Physikalisches Institut, Universit\"{a}t Bern, Sidlerstrasse 5, 3012 Bern, Switzerland}}

\email{david.berardo@mcgill.ca}
\email{andrew.cumming@mcgill.ca}
\email{gabriel.marleau@space.unibe.ch}

\newcommand{\update}[1]{\textcolor{black}{#1}}

\begin{abstract}
We calculate the evolution of gas giant planets during the runaway gas accretion phase of formation, to understand how the luminosity of young giant planets depends on the accretion conditions. We construct steady-state envelope models, and run time-dependent simulations of accreting planets with the Modules for Experiments in Stellar Astrophysics (MESA) code. We show that the evolution of the internal entropy depends on the contrast between the internal adiabat and the entropy of the accreted material, parametrized by the shock temperature $T_0$ and pressure $P_0$. At low temperatures ($T_0\lesssim 300$--$1000\ {\rm K}$, depending on model parameters), the accreted material has a lower entropy than the interior. The convection zone extends to the surface and can drive a large luminosity, leading to rapid cooling and cold starts. For higher temperatures, the accreted material has a larger entropy than the interior, giving a radiative zone that stalls cooling. For $T_0\gtrsim 2000\ {\rm K}$, the surface--interior entropy contrast cannot be accommodated by the radiative envelope, and the accreted matter accumulates with high entropy, forming a hot start. The final state of the planet depends on the shock temperature, accretion rate, and starting entropy at the onset of runaway accretion. Cold starts with $L\lesssim 5\times 10^{-6}\ L_\odot$ require low accretion rates and starting entropy, and that the temperature of the accreting material is maintained close to the nebula temperature. If instead the temperature is near the value required to radiate the accretion luminosity, $4\pi R^2\sigma T_0^4\sim (GM\dot M/R)$, as suggested by previous work on radiative shocks in the context of star formation, gas giant planets form in a hot start with $L\sim 10^{-4}\ L_\odot$.
\end{abstract}



\section{Introduction}
\label{sec: Intro}

The direct detection of young gas giant planets is an important test of planet formation mechanisms, because at young ages the planet has had less time to thermally relax and so its thermal state depends on how it formed \citep{Stevenson1982,Fortney2005,Marley2007,Fortney2008}. Traditional cooling models for brown dwarfs and giant planets were based on hot initial (post-formation) conditions, in which case the thermal time is short and the planet quickly forgets the initial conditions and evolves onto a cooling track that depends only on the mass (e.g.~\citealt{Burrows1997,Baraffe2003}). \cite{Fortney2005} and \cite{Marley2007} pointed out that gas giants formed by core accretion might be much colder than these earlier ``hot start'' models. They showed that the core accretion model described in the series of papers \cite{Pollack1996}, \cite{Bodenheimer2000}, and \cite{Hubickyj2005} produced planets that were significantly less luminous, implying that giant planets instead have a ``cold start''. 

Given uncertainties in planet formation models and the potential large range in luminosity of newly formed gas giant planets, \cite{Spiegel2012} took the approach of treating the internal entropy of the gas giant after formation as a free parameter, producing a range of ``warm starts''. The predicted cooling tracks then depend on the planet mass and initial entropy. \cite{Bonnefoy2013} and \cite{Marleau2014} explored the joint constraint on these two parameters that can be inferred from a directly imaged planet with a known luminosity and age. For hot initial conditions, the cooling tracks depend only on the mass; cold initial conditions require a more massive planet to match the observed luminosity. Fitting hot start cooling curves therefore gives a lower limit on the planet mass. Matching the observed luminosity gives a lower limit on the initial entropy, because of the sensitive dependence of luminosity on the internal entropy (e.g.~fig.~2 of \citealt{Marleau2014}). Additional information about the planet mass, such as an upper limit from dynamics, can break the degeneracy and reduce the allowed range of initial entropy. 

The population of directly-imaged planets shows a wide range of luminosity (e.g.~\citealt{Neuhauser2012,Bowler2016}), with most being too luminous to be cold starts. Examples are $\beta$ Pic b with $L\approx 2\times 10^{-4}\ L_\odot$ \citep{Lagrange2009,Lagrange2010,Bonnefoy2013}, or the HR8799 planets with $L\approx 2\times 10^{-5}\ L_\odot$ for HR8799c, d, and e, and $8\times 10^{-6}\ L_\odot$ for HR8799b \citep{Marois2008,Marois2010}. The inferred initial entropies in these cases are significantly larger than in \cite{Marley2007} (\citealt{Bonnefoy2013,Bowler2013,Currie2013,Marleau2014}). The best case for a cold start is the young giant planet 51 Eri b,
which has a projected separation of 13~au from its star and $L\approx 1.4$--$4\times 10^{-6}\ L_\odot$ \citep{Macintosh2015}. This luminosity is consistent with the value $\approx 2\times 10^{-6}\ L_\odot$ predicted by \citep{Marley2007}, but it also matches a hot start for a planet mass $2$--$3\ M_J$ at the stellar age $\approx 20\ {\rm Myr}$. 
Similarly, the low effective temperature of  $850\ {\rm K}$ for HD 131399Ab corresponds to a hot start mass of $4\ M_J$ at $16\ {\rm Myr}$ \citep{Wagner2016}. Another cold object is GJ~504b, which has an effective temperature of only $510\ {\rm K}$ 
\citep{Kuzuhara2013}, but indications that the star is Gyrs old imply that it may be a low-mass brown dwarf rather than a planet \citep{Fuhrmann2015,DOrazi2016}. 

Interesting from the point of view of testing formation models has been the discovery of protoplanets still embedded in a protoplanetary disk. For example, HD~100546 b is a directly-imaged object 50~au from its Herbig Ae/Be host with a luminosity $\sim 10^{-4}\ L_\odot$ \citep{Quanz2013,Currie2014a,Quanz2015}, and the star may host a second planet closer in \citep{Currie2015,Garufi2016}. 
\cite{Sallum2015} identified two and perhaps three accreting protoplanets in the LkCa~15 transition disk. The infrared and H\,$\alpha$ luminosities were consistent with expected accretion rates:
\citet{Sallum2015} report $M\dot M\sim 10^{-5}\ M_J^2\ {\rm yr}^{-1}$,
where $M$ and $\dot M$ are respectively the planetary mass and accretion rate,
which agrees with typical accretion rates of $\sim 10^{-3}$--$10^{-2}\ M_\oplus\ {\rm yr^{-1}}$ in models (e.g.~\citealt{Lissauer2009}) for $M\sim M_J$. 
The young ages of these stars $\lesssim 10\ {\rm Myr}$ correspond to early times when there is greater potential for distinguishing formation models (e.g.~fig.~4 of \citealt{Marley2007}), \update{especially since the planets could be substantially younger than the star \citep{Fortney2005}.} The interpretation of the observations is complicated, however. Contributions from the environment around the protoplanet, which is likely still accreting, need to be considered, and if accretion is ongoing the accretion luminosity $L_{\rm accr}\approx GM\dot M/R$, where $R$ is the planetary radius, may dominate the internal luminosity. \update{Nevertheless, these effects can potentially be distinguished by studying the spectral energy distribution or spatially resolving the emission. For example, observations of HD~100546 b are able to make out a point-source component (surrounded by spatially-resolved emission) with blackbody radius and luminosity consistent with those of a young gas giant \citep{Currie2014b,Quanz2015}.}

Interpreting the current and upcoming observations of young gas giants requires understanding more fully the physics that sets the thermal state of the planet during and immediately after formation. \cite{Marley2007} emphasized that because most of the mass of the gas giant is delivered through an accretion shock, the efficiency with which the shock radiates away the gravitational energy of the accreted matter is a key uncertainty, determining the temperature of the material added to the planet by accretion. The need to accurately treat the radiative cooling at the shock (in particular whether the shock is supercritical, e.g.~see \citealt{Commercon2011}) has been discussed in
\textsection~8.1 of \citet{Mordasini2012} and in
reviews such as \cite{Chabrier2014}. \cite{Mordasini2013} also identified the planetesimal surface density in the disk as a key ingredient since it sets the core mass. He simulated the growth of planets under cold- and hot-start conditions by changing the outer boundary condition for the planet during the accretion phase. In the cold case, the final entropy of the planet was found to depend sensitively on the resulting core mass through the feedback action of the accretion shock. Most recently, \cite{Owen2016} pointed out the potential importance of non-spherical accretion and studied the role of an accretion boundary layer in setting the thermal state of the accreted matter.

In this paper, we focus on the phase of the core accretion scenario in which the accreting matter forms a shock at the surface of the planet. This runaway accretion phase occurs once the contraction rate of the gas envelope surrounding a newly formed core of $\sim 10\ M_\oplus$ becomes larger than the rate at which the disk can supply mass to the envelope (e.g.~\citealt{Helled2014,Mordasini2015}). The planet then shrinks within its Hill sphere and mass flows hydrodynamically onto the planet. Given the uncertainty in the temperature of the post-shock material, we treat the entropy at the surface of the planet as a free parameter. The aim is to better understand how the matter deposited by the accretion shock becomes part of the planet and therefore sets the internal entropy. This approach is similar to previous work on accreting protostars in which the efficiency of the accretion shock is treated as a free parameter (e.g.~\citealt{Prialnik1985,Siess1997,Baraffe2009}; see discussion in \S~\ref{sec:previoushotcold}). We improve on the previous calculations of core accretion with hot outer boundaries by \cite{Mordasini2012} and \cite{Mordasini2013}, which assumed constant luminosity inside the planet and only global energy conservation, by following the full internal energy profile during accretion.

\begin{figure}
\epsscale{1.2}
\plotone{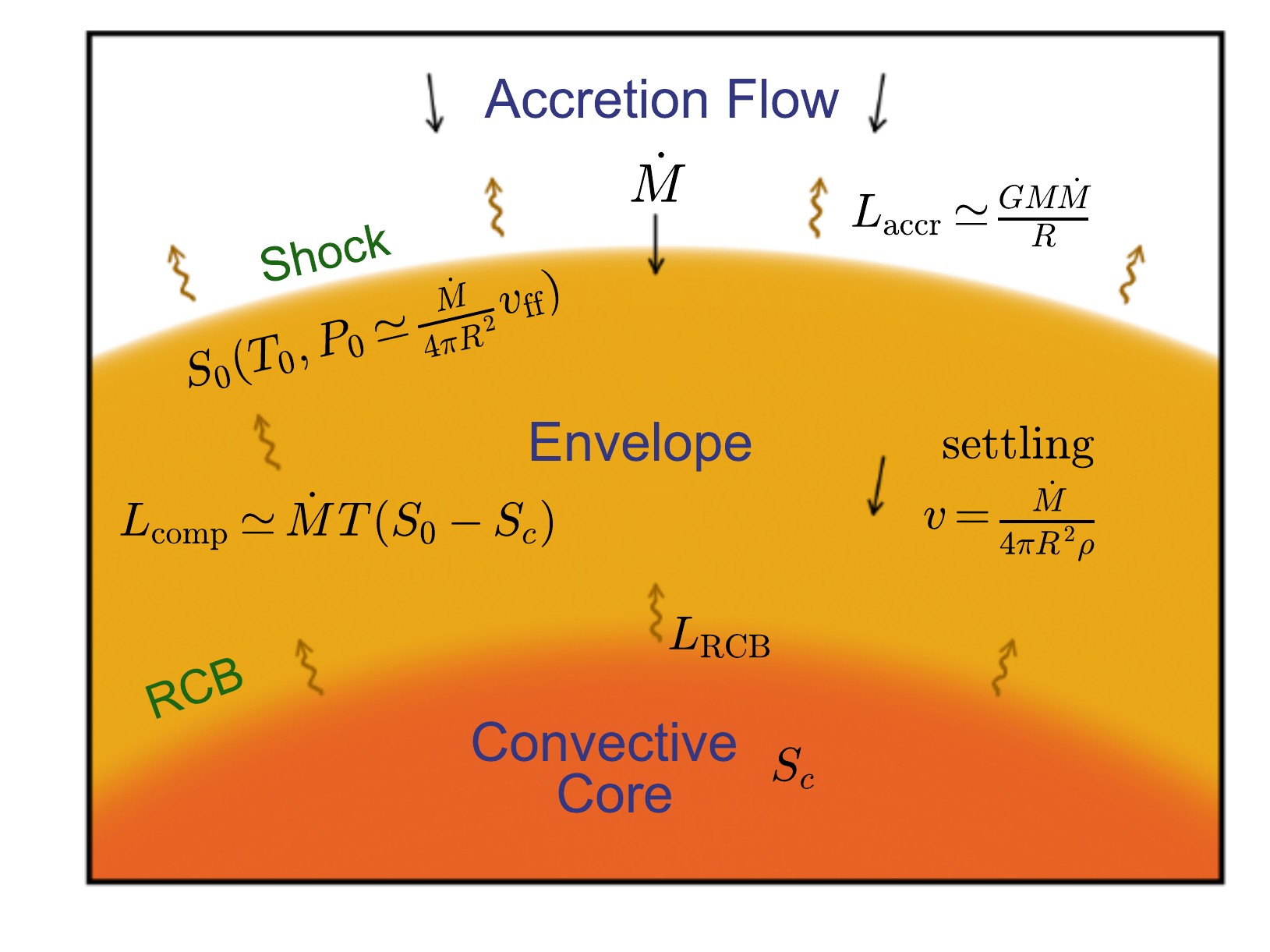}
\caption{Diagram of a spherically-symmetrically accreting gas giant. Shown are the last parts of the
accretion flow (\textit{top}), the radiative
envelope (\textit{middle}), and the convective interior (\textit{bottom}).
Matter accretes onto the envelope with a rate $\dot{M}$, where it shocks and releases energy as an accretion luminosity $L_\mathrm{accr}$. Immediately after the shock, the matter has temperature $T_0$, pressure $P_0$ equal to the ram pressure (eq.~[\ref{eq:Paccr}]), and thus entropy $S_0$. As the material settles down through the envelope to the convective core with a velocity $v=\dot M/4\pi r^2\rho$, it releases an additional luminosity $L_\mathrm{comp}$ from compressional heating and finally reaches the radiative-convective boundary (RCB). The convective core has entropy $S_c$ and supplies a luminosity $L_\mathrm{RCB}$ to the base of the envelope.}
\label{fig:schematic}
\end{figure}

A schematic diagram of the different regions we consider in this paper is shown in Figure \ref{fig:schematic}. We start in \S~\ref{sec:entropy} by discussing the expected values of entropy of the accreted material deposited by the accretion shock at the surface of the planet. In \S~\ref{sec:theorymodels} we compute thermal steady state models of the accreting envelope to understand how freshly accreted material becomes part of the planet, following \cite{Stahler1988} who studied the envelopes of accreting low-mass protostars. We show that there are three regimes of accretion depending on how the entropy of the newly accreted material compares to the internal adiabat. In \S~\ref{sec:mesamodels}, we numerically calculate the evolution of gas giants accreting matter with a range of entropy, using the Modules for Experiments in Stellar Astrophysics (MESA) code \citep{Paxton2011,Paxton2013,Paxton2015}, and investigate the sensitivity of the final thermal state of the planet to the shock conditions and starting entropy at the onset of accretion. We summarize, compare our results to observed systems, and discuss the implications in \S~\ref{sec:summdisc}. Finally, analytical formul\ae\ for the entropy of an ideal gas as well as analytic solutions of envelope structures of accreting atmospheres are presented in Appendices~\ref{append A} and~\ref{append B} respectively.

\section{Entropy of the post-shock gas}
\label{sec:entropy}

In this section, we discuss the state of the gas just after the accretion shock. 

\subsection{Previous approaches to hot and cold accretion}
\label{sec:previoushotcold}

There have been a few different approaches in the literature to modelling the unknown radiative efficiency of the accretion shock in accreting protostars and planets. This results in different assumptions about the post-shock temperature and entropy ($S_0$ and $T_0$ in Fig.~\ref{fig:schematic}).

In the context of gas giant formation, the core accretion models of \cite{Pollack1996}, \cite{Bodenheimer2000}, and \cite{Hubickyj2005} are based on the assumption that the shock is isothermal, with a temperature set by integrating the radiative diffusion equation inwards through the spherical accretion flow from the nebula (i.e.~the local circumstellar disk) to the shock. In the limit where the flow is optically thin, the shock temperature is then the nebula temperature, but could be much larger if the flow is optically thick (see discussion in \S~2 of \citealt{Bodenheimer2000}). The cold accretion limit of these models is therefore that the post-shock temperature of the gas is $T_0=T_{\rm neb}$, or $150\ {\rm K}$ in the calculations of \cite{Hubickyj2005} (although whether the temperatures in the models corresponding to the \citealt{Marley2007} cold starts were that low was not explicitly reported).

An alternative approach that has been used in a variety of contexts is to model the shock efficiency by the fraction of the specific accretion energy $GM/R$ that is incorporated into the star or planet. This is implemented either by adding an amount $\alpha GM/R$ to the specific internal energy of the accreted matter if following the detailed structure with a stellar evolution code (\citealt{Prialnik1985,Siess1997,Baraffe2009}), or by adding a contribution $\alpha GM\dot M/R$ to the luminosity if following the global energetics (\citealt{Hartmann1997}). For gas giant accretion, \cite{Mordasini2012} and \cite{Mordasini2013} step through sequences of detailed planet models by tracking the global energetics, and model cold or hot accretion by not including or including the accretion luminosity in the internal luminosity of the planet. \cite{Owen2016} recently applied the approach of \cite{Hartmann1997} to disk-fed planetary growth, calculating $\alpha$ as set by the disk boundary layer.

In these approaches, the cold limit corresponds to setting $\alpha=0$, which means that the accreting material adjusts its temperature to match the gas already at the surface. With this boundary condition, the cooling history of the accreting object is affected by accretion only through the fact that its mass is growing, which changes its thermal timescale. Even for $\alpha=0$, the temperature at the surface can be much larger than $T_{\rm neb}$, and so this is a different cold limit than in \cite{Bodenheimer2000}. For example, taking a typical internal luminosity $L_{\rm int}\sim 10^{-4}\  L_\odot$ and planet radius $2\ R_J$ gives $T_0 = T_{\rm therm} \approx (L_{\rm int}/4\pi R^2\sigma)^{1/4}\approx 1300\ {\rm K}$, where $\sigma$ is the Stefan--Boltzmann constant.

In the hot limit with $\alpha=1$, the surface temperature is given by $T_0=T_{\rm hot}\approx (L_{\rm accr}/4\pi R^2\sigma)^{1/4}$ where $L_{\rm accr}\approx GM\dot M/R$ is the accretion luminosity, 
\begin{equation}
L_{\rm accr}\approx4.4\times 10^{-3}\ L_\odot\ \left({\dot M\over 10^{-2}\ M_\oplus\,{\rm yr^{-1}}}\right)\left({M\over M_J}\right)\left({R \over 2\ R_J}\right)^{-1}.
\end{equation}
This gives a temperature
\begin{equation}\label{eq:hotT}
T_{\rm hot} \approx 3300\ {\rm K}\ \left({\dot M\over 10^{-2}\ M_\oplus\,{\rm yr^{-1}}}\right)^{1/4}\left({M\over M_J}\right)^{1/4}\left({R \over 2\ R_J}\right)^{-3/4}.
\end{equation}
We have scaled to a typical accretion rate during the runaway accretion phase of $\dot M\lesssim 10^{-2}\ M_\Earth\ {\rm yr^{-1}}=1.9\times 10^{18}\ {\rm g\ s^{-1}}$ \citep{Pollack1996,Lissauer2009} and use as everywhere $R_J=7.15\times10^9$~cm.

Shock models suggest that the post-shock temperature is more likely to be close to $T_{\rm hot}$ than $T_{\rm neb}$. \cite{Stahler1980} argued that, even if the accretion flow is optically thin, the outer layers of the protostar (or here the planet) will be heated because some of the energy released in the shock is radiated inwards (see fig.~5 of \citealt{Stahler1980} and associated discussion; see also the discussion in \citealt{Calvet1998} and \citealt{Commercon2011}). For an optically thin accretion flow, \cite{Stahler1980} derived the relation $4\pi R^2\sigma T^4\approx (3/4)L_{\rm accr}$ for the post-shock temperature (see their eq.~[24]), which is $(3/4)^{1/4}T_{\rm hot}\approx 3100\ {\rm K}$. The factor of $3/4$ relies on an approximate estimate of the outwards radiation that is reprocessed and travels back inwards towards the surface, but the temperature is only weakly affected (for example a factor 1/4 would still give $2300\ {\rm K}$). This suggests that the temperature in the post-shock layers is $T_0\gg T_{\rm neb}$ and even $T_0\gg T_{\rm therm}$. However, since detailed calculations of the radiative transfer associated with the shock are in the early stages (e.g.\ \citealt{Marleau2016}), we will treat $T_0$ as a free parameter and consider values in the full range from $\approx T_{\rm neb}$ to $\approx T_{\rm hot}$.

\subsection{The physical conditions post-shock}

We now discuss the conditions post-shock, taking the temperature $T_0$ as a parameter. Following \cite{Bodenheimer2000}, we consider an isothermal shock with density jump  $\rho_2/\rho_1=v_{\rm ff}^2/c_s^2$, where the matter arrives at the free fall velocity $v_{\rm ff} = (2GM/R)^{1/2} = 42\ {\rm km\ s^{-1}}\ (M/M_J)^{1/2}(2\ R_J/R)^{1/2}$, and $c_s$ is the isothermal sound speed. The post-shock pressure is the ram pressure $P_{\rm accr}=\rho_2 c_s^2={\dot M v_{\rm ff}/4\pi R^2}$ or
\begin{eqnarray}\label{eq:Paccr}
P_{\rm accr} &=& 3.1\times 10^3\ {\rm erg\ cm^{-3}}\ \left({\dot M\over 10^{-2}\ M_\oplus\,{\rm yr^{-1}}}\right)\nonumber\\
&&\times \left({M\over M_J}\right)^{1/2}\left({R\over 2\ R_J}\right)^{-5/2}.
\end{eqnarray}
(Note that in this paper, we use cgs units for pressure; recall that $P=1\ {\rm bar}=10^6\ {\rm erg\ cm^{-3}}$.)

At the low densities near the surface of the planet, the equation of state is close to an ideal gas. In Appendix~\ref{append A} we show that for a mixture of H$_2$ and He with helium mass fraction $Y=0.243$ (matching the value used by \citealt{Pollack1996}) the entropy\footnote{\update{Throughout this work, entropies have the same reference point as the published tables of \citet{Saumon1995}, and hence can be compared directly to the MESA code and \citet{Marleau2014}. When comparing to other works, it is important to note that a different reference point may have been chosen (see fig.~4 and appendix~B of \citealt{Marleau2014}).}}
per baryon is
\begin{equation}\label{eq:Sformula}
{S\over k_{\rm B}/m_p}\approx 10.8 + 3.4\log_{10} T_3 - 1.0\log_{10} P_4,
\end{equation}
where $k_{\rm B}$ is Boltzmann's constant, $m_p$ is the proton mass, and $T_3 \equiv T/(1000\ {\rm K})$, $P_4\equiv P/(10^4\ {\rm erg\ cm^{-3}})$. Using the ram pressure (eq.~[\ref{eq:Paccr}]) and assuming the gas remains molecular post-shock, the post-shock entropy $S_0$ is therefore
\begin{eqnarray}\label{eq:coldstartS}
{S_0\over k_{\rm B}/m_p} &\approx& 7.4 - \log_{10}\left({\dot M\over 10^{-2}\ M_\oplus\,{\rm yr^{-1}}}\right)+3.4\,\log_{10}\left({T_0\over 150\ {\rm K}}\right)\nonumber\\
&&-0.51\,\log_{10}\left({M\over M_J}\right)+2.5\,\log_{10}\left({R\over 2\ R_J}\right),\end{eqnarray}
where we have scaled to the lowest possible temperature expected for $T_0$, the nebula temperature in \cite{Hubickyj2005}. At higher temperatures, the hydrogen will be atomic post-shock, in which case the entropy is (Appendix~\ref{append A})
\begin{equation}\label{eq:Sformula2}
{S\over k_{\rm B}/m_p}\approx 17.2 + 4.7\log_{10} T_3 - 1.9\log_{10} P_4.
\end{equation}
The maximal value of entropy we expect is for $T_0\approx T_{\rm hot}$ (eq.~[\ref{eq:hotT}]), which gives 
\begin{eqnarray}\label{eq:hotstartS}
{S_0\over k_{\rm B}/m_p} &\approx & 20.6 - 0.72\,\log_{10}\left({\dot M\over 10^{-2}\ M_\oplus\,{\rm yr^{-1}}}\right)\nonumber\\&&+0.23\,\log_{10}\left({M\over M_J}\right)+1.17\,\log_{10}\left({R\over 2 R_J}\right).
\end{eqnarray}
We see that there is a large variation in $S_0$, the entropy of the material deposited at the planet surface, depending on the shock temperature. These values can be larger or smaller than the internal entropy of the planet at the moment runaway accretion begins (which for example is $S\approx 11\ k_{\rm B}/m_p$ in the simulations of \citealt{Mordasini2013}). In the next section we investigate the response of the planet to accretion in these different cases.


\section{The structure of the accreting envelope}
\label{sec:theorymodels}

To understand the evolution of the accreting gas after arrival on the planet, we first construct envelope models following the approach of \cite{Stahler1980} and \cite{Stahler1988} for accreting low-mass protostars. In the envelope, the entropy profile adjusts from the surface value $S_0$ to the interior value $S_c$. The thermal timescale across the envelope is short compared to the evolution time, so that we can assume thermal equilibrium for the envelope. Indeed, we will see in the time-dependent simulations in the next section that the envelope adopts a self-similar profile, slowly adjusting over longer timescales as the internal entropy changes.

\subsection{Envelope models}
\label{sec:envmodels}

We follow \cite{Stahler1988} and construct a plane-parallel envelope model in thermal equilibrium with constant gravity $g=GM/R^2=6.2\times 10^2\ {\rm cm\ s^{-2}} (M/M_J)(R/2\ R_J)^{-2}$. This is a good approximation since the envelope is thin: $H_P/R\approx 0.005\ (T/1000\ {\rm K}) (R/2\ R_J)(M/M_J)^{-1}(\mu/2)^{-1}$ where $H_P=k_{\rm B}T/\mu m_pg$ is the pressure scale height and $\mu$ the mean molecular weight. The entropy equation is
\begin{equation} \label{eq:entropy1}
  T{\partial S\over \partial t}+vT{\partial S\over \partial r}=-{1\over 4\pi r^2\rho}{\partial L\over \partial r},
\end{equation}
where $L(r)$ is the luminosity at radius $r$. Mass continuity gives the velocity of the settling material $v = -\dot M/4\pi r^2 \rho$. Switching to pressure as an independent coordinate using hydrostatic balance $dP/dr =-\rho g$, and assuming a steady state, equation (\ref{eq:entropy1}) becomes
\begin{equation}\label{eq:entropy2}
  \dot M T{dS\over dP}={dL\over dP}.
\end{equation}
As pointed out by \cite{Stahler1988}, this shows that, to the extent that temperature is constant, $L-\dot M TS$ is constant in the envelope, so that in particular the change in luminosity $\Delta L$ across the envelope is related to the change in entropy $\Delta S$ as $\Delta L\approx \dot M T \Delta S$.

To calculate the envelope models, we rewrite equation (\ref{eq:entropy2}), and integrate equations for $T$ and $L$ as a function of pressure,
\begin{eqnarray}
{dT\over dP}&=&{T\over P}\nabla\\
{dL\over dP}&=&\dot M T\left[\left.{\partial S\over \partial P}\right|_T + \left.{\partial S\over \partial T}\right|_P {dT\over dP}\right],
\end{eqnarray}
where $\nabla=d\ln T/d\ln P$ is the temperature gradient in the planet\footnote{The code used to calculate the envelope models is available at \url{https://github.com/andrewcumming/gasgiant}.}. We use the equation of state tables from the MESA code for our integrations and assume the composition of the atmosphere is hydrogen and helium with helium mass fraction $Y=0.243$ \citep{Pollack1996}. We integrate inwards to a pressure of $10^8\ {\rm erg\ cm^{-3}}$ where the density is typically $\sim 3\times 10^{-4}\ {\rm g\ cm^{-3}}$. Under these conditions the equation of state is close to an ideal gas (e.g.~see Fig.~1 of \citealt{Saumon1995}), and we find similar results assuming an ideal gas equation of state and calculating the dissociation fraction of the molecular hydrogen using the Saha equation as outlined in Appendix~\ref{append A}. The mass and radius of the planet are free parameters in the envelope model. We use the giant planet models of \cite{Marleau2014} to self-consistently determine the radius corresponding to the internal entropy of the planet, matching the entropy of the convection zone at the base of the envelope model.

The temperature gradient $\nabla$ depends on the heat transport mechanism. For radiative diffusion,  $\nabla=\nabla_{\rm rad}$ given by the radiative diffusion equation
\begin{equation}\label{eq:raddiffusion}
L = -4\pi r^2 {4acT^3\over 3\kappa\rho}{dT\over dr} =  {16\pi acT^4GM\over 3\kappa P}\nabla_{\rm rad}.
\end{equation}
We calculate the opacity $\kappa$ using the tables supplied with MESA, choosing the low-temperature tables based on \cite{Freedman2008,Freedman2014} with $Z=0.02$ (the \texttt{lowT\_Freedman11\_z0.02.data} table). These opacities do not include grain opacity, which is significantly uncertain because small grains may coagulate and settle out of the atmosphere \citep{Podolak2003,Movshovitz2008}. Core accretion models often assume a fixed grain contribution, e.g.~2\% of interstellar values \cite{Pollack1996}. \cite{Movshovitz2010} modelled grain evolution up to crossover mass and found that the grain opacity was even lower. \cite{Mordasini2014a} and \cite{Mordasini2014b} compared planet population synthesis models with observations, preferring a grain opacity of 0.3\% of the interstellar value. In most of the models in this paper, we include only the gas opacity and assume that grain opacity is not significant. We investigate the influence of grain opacity in \S~\ref{sec:grains}.

The post-shock material is typically in the free-streaming regime, i.e.~it is optically thin over a few post-shock pressure scale heights. Indeed,
defining the photosphere to be where the optical depth as measured from the shock $\Delta\tau\approx 1$, the photospheric pressure $P_{\rm phot}\approx g/\kappa$
is larger than the ram pressure $P_{\rm accr}$ by a factor of $f$ times $e\approx2.7$ when $\kappa\lesssim 0.02\ {\rm cm^2\ g^{-1}}\ (f/3)^{-1}(\dot M/0.01\ M_\oplus\ {\rm yr^{-1}})(RM/2R_JM_J)^{1/2}$ (see eq.~[\ref{eq:Paccr}]), which is generally satisfied when the grain contribution to the opacity is suppressed to the percent level.
Note that $P_{\rm phot}\sim g/\kappa$ holds regardless of the optical thickness of the upstream accretion flow since the post-shock gas is (nearly) in hydrostatic equilibrium, which implies $P\sim\rho\Delta r g$ where $\Delta r$ is the distance from the shock.

Equation (\ref{eq:raddiffusion}) remains valid in free-streaming conditions,
under the assumptions of a grey opacity, local thermodynamic equilibrium, and the Eddington approximation  (e.g.~\citealt{Hubeny2014}).
We do not follow energy deposited within the (optically-thin) outer layers of the envelope due to irradiation by the accretion shock. Instead, we
include the influence of the accretion shock by setting the temperature $T_0$ at the post-shock ram pressure $P=P_{\rm accr}$.
This approach should be valid but could be verified by a detailed calculation of the radiative transfer through the shock and in the outer layers.

When $\nabla_{\rm rad}>\nabla_{\rm ad}$, where $\nabla_{\rm ad}=(\partial \ln T/\partial \ln P)_S$ is the adiabatic gradient, convection transports energy. In that case, we calculate $\nabla$ from mixing length theory following \cite{Henyey1965} (see p.~558 of \citealt{Hubeny2014} for a useful summary). For efficient convection, the convective luminosity is
\begin{equation}\label{eq:Lconv}
L_{\rm conv} = 4\pi R^2 {1\over 2}\rho v_{\rm conv} c_P T \left(\nabla-\nabla_{\rm ad}\right),
\end{equation}
where we set the mixing length equal to the pressure scale height, $c_P$ is the heat capacity per unit mass at constant pressure, and the convective velocity is $v_{\rm conv}\approx (gH/8)^{1/2}(\nabla-\nabla_{\rm ad})^{1/2}=(P/8\rho)^{1/2}(\nabla-\nabla_{\rm ad})^{1/2}$. Near the surface of the convection zone, the $\nabla-\nabla_{\rm ad}$ term can be of order unity. The convection extends into optically thin ($\Delta\tau\ll 1$) regions of the envelope for low shock temperatures, and radiative losses from convective elements reduce the convective efficiency. We account for this using the prescription of \cite{Henyey1965} using $\Delta\tau$ as the optical depth. It is not clear whether this applies for the situation of a bounded atmosphere irradiated by the accretion shock and in which the accretion flow above the shock can be optically thick. However, we find that including radiative losses in the mixing length prescription changes the luminosity in the envelope by less than a few percent.

\subsection{Structure of the envelope for different boundary temperatures}
\label{sec:envelope models}

\begin{figure}
\epsscale{1.2}
\plotone{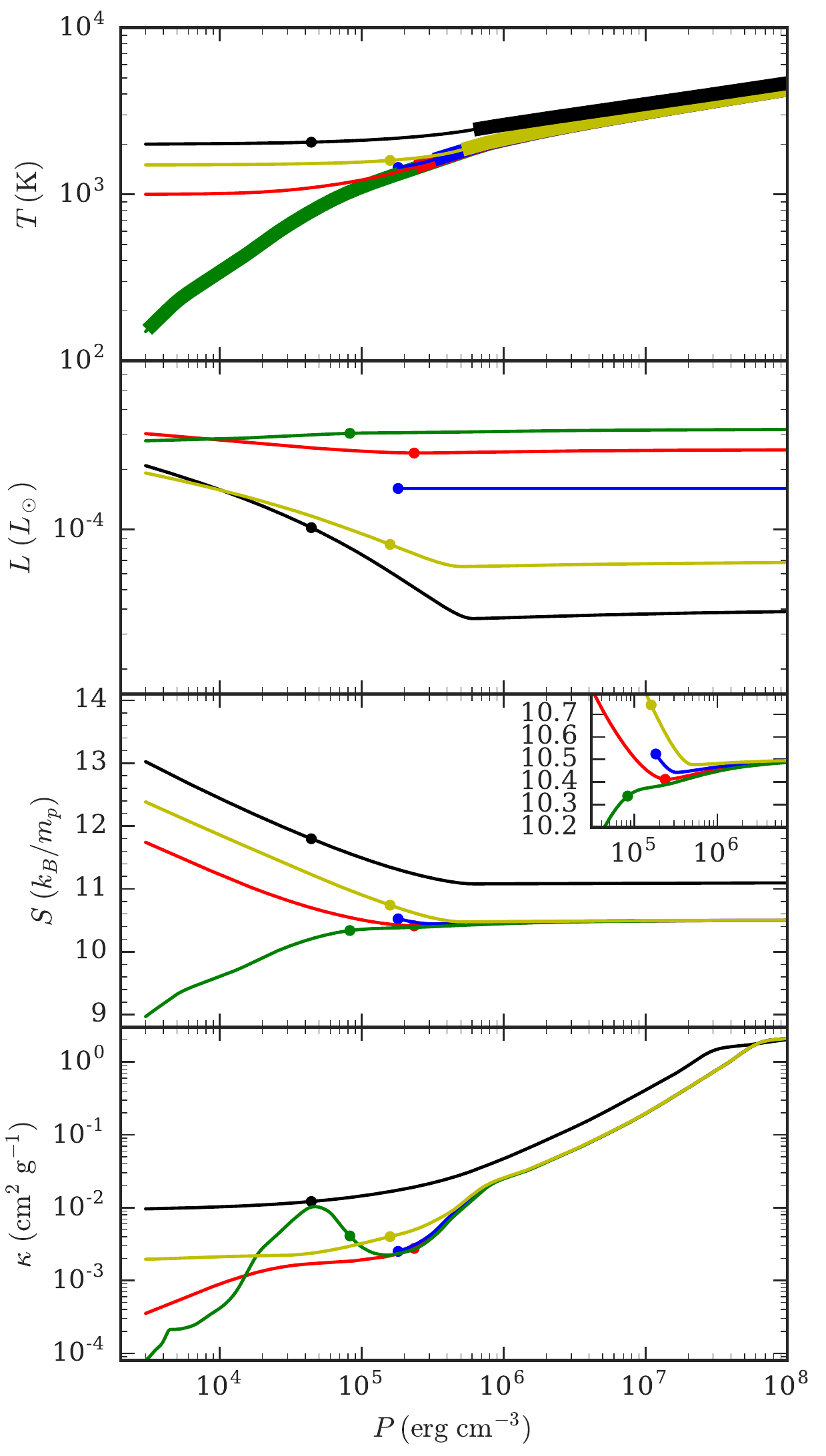}
\caption{Envelope profile for different choices of outer boundary temperature. The black, yellow, red, and green curves are for outer temperatures $T_0=2000, 1500, 1000,$ and $150\ {\rm K}$ at a pressure $P_0=3\times 10^3\ {\rm erg\ cm^{-3}}$. Except for the hottest model, we have chosen the luminosity of the different envelope models so that they match onto a convection zone with entropy $10.5\ k_{\rm B}/m_p$ at depth. Blue is for a cooling boundary condition with no accretion. In all cases, the planet has mass $1\,M_J$ and radius $2\,R_J$.  The accretion rate for the accreting envelopes is $\dot M=0.01\ \dot M_{\Earth}\ {\rm yr^{-1}}$. The filled circles show the location where the optical depth from the shock $\Delta\tau=2/3$. The region of convection is indicated by thick lines in the temperature profiles in the upper panel. The inset shows the region near the radiative-convective boundary.}
\label{fig:profile}
\end{figure}

Figure \ref{fig:profile} shows example profiles of the accreting envelope for the same accretion rate and internal adiabat, but with different outer boundary temperatures. We model a $1\ M_J$, $2\ R_J$ planet accreting at $0.01\ M_\Earth\ {\rm yr^{-1}}$. We adjust the luminosity at the top of the atmosphere to try to match the entropy at the base of the atmosphere at $P=10^8\ {\rm erg\ cm^{-3}}$ to $S_c=10.5\ {k_{\rm B}/m_p}$, the appropriate value of internal entropy for $2\ R_J$ (e.g.~Fig.~A1 of \citealt{Marleau2014}). The outer boundary is placed at the ram pressure which is $3\times 10^3\ {\rm erg\ cm^{-3}}$ from equation (\ref{eq:Paccr}). We also show the envelope profile for an isolated, non-accreting planet for comparison, where we set the outer pressure to $P_{\rm phot}=(2/3)(g/\kappa)$ and set the temperature to $T_{\rm eff}=(L/4\pi R^2\sigma)^{1/4}$. 

We find that the structure and luminosity of the accreting envelope depends on the entropy at the outer boundary. If the surface entropy is significantly larger than the internal entropy, the radiative-convective boundary is pushed deeper, and the luminosity there, $L_{\rm RCB}$, is smaller. This is important because $L_{\rm RCB}$ determines how quickly the convective core cools down and moves to lower entropy. The effect of the hot envelope is therefore to reduce the cooling luminosity and increase the cooling timescale of the planet. At lower surface entropy, the material in the envelope reaches lower entropy than the convection zone. This entropy inversion enhances convection, moving the RCB outwards and increasing the cooling luminosity. The models with hotter outer temperatures of $1500$ and $2000\ {\rm K}$ in Figure \ref{fig:profile} are examples of envelopes with reduced cooling luminosity.

\label{sec:stalling boundary model}

The entropy and luminosity profiles in the envelope are similar to those considered by \cite{Stahler1988} (see fig.~4 of that paper). 
The luminosity 
increases outwards due to compressional heating, which supplies a luminosity $L_{\rm comp}\approx\dot M T\Delta S$ or 
\begin{eqnarray}
  L_{\rm comp}&\approx & 8\times 10^{-5}\ L_\odot\ \left({\dot M\over 10^{-2}\ M_\oplus\,{\rm yr^{-1}}}\right)\nonumber\\
                & & \times\left({T\over 2000\ {\rm K}}\right)\left({\Delta S\over k_{\rm B}/m_p}\right).
\end{eqnarray}
The entropy decreases inwards in the radiative zone, joining smoothly onto the convection zone at the radiative-convective boundary (RCB). In the convection zone, the entropy initially increases slightly inwards and then levels off as convection becomes efficient and dominates the energy transport. The hot outer boundary pushes the RCB deeper into the planet than in a non-accreting planet with the same internal entropy. This makes the luminosity leaving the convective core smaller (e.g.~\citealt{Burrows2000,Arras2006}), so that the core cools more slowly. For the $T_0=2000\ {\rm K}$ case, the RCB moves inwards by about a factor of 2 in pressure, and the cooling is slower by about a factor of 4 relative to a non-accreting planet.

\label{sec:cooling boundary model}

In the colder model with an outer temperature of $1000\ {\rm K}$, the entropy quickly drops below the entropy of the convection zone on moving inwards through the envelope. Convection extends out almost to the photosphere, and the luminosity is larger than in the non-accreting case. The potential for enhanced luminosity can be understood by considering the entropy gradient in the planet, which is \citep{Stahler1988}
\begin{equation}
  {dS\over d\ln P} = c_P \left(\nabla-\nabla_{\rm ad}\right).
\end{equation}
When $c_P = (7/4)(k_{\rm B}/m_p)$ (assuming pure H$_2$ with $\mu=2$ and only translational and rotational degrees of freedom) a change in entropy $\Delta S$ across a pressure range $\Delta \log_{10}\, P$ implies
\begin{equation}
   \nabla-\nabla_{\rm ad}\approx 0.25\,\left({\Delta S\over k_{\rm B}/m_p}\right){1\over \Delta \log_{10}\,P}.
\end{equation}
This significant departure from adiabaticity near the outer boundary is needed to increase the entropy from its value at the outer edge of the convection zone to the value at the center, $S_c$. From equation (\ref{eq:Lconv}), the luminosity resulting from this superadiabaticity is
\begin{eqnarray}
  L_{\rm conv} &\sim& 10^{-3}\ L_\odot\ \left({R\over 2\,R_J}\right)^2\left({\nabla-\nabla_{\rm ad}\over 0.25}\right)^{3/2}\nonumber\\&&
 \left({T \over 1000~{\rm K}}\right)^{1/2}\left({P\over 10^5\ {\rm erg\ cm^{-3}}}\right).
\end{eqnarray}
The luminosities of the envelopes with the colder outer boundaries are therefore greater than the cooling luminosity of the planet without accretion, by a factor of 1.5 for $T_0=1000\ {\rm K}$, and a factor of two for $T_0=150\ {\rm K}$, which is convective all the way out to the outer boundary.

\subsection{Hot accretion: a minimum luminosity and minimum entropy for hot envelopes}

\label{sec:hot boundary model}

The hottest model in Figure \ref{fig:profile}, with $T_0=2000\ {\rm K}$, does not match onto an internal adiabat with $S_c=10.5\ k_{\rm B}/m_p$. Constructing envelope models with different luminosities, 
the lowest entropy that we can match onto with an outwards luminosity is $S_c=11.1\ k_{\rm B}/m_p$, which is the model shown in Figure \ref{fig:profile}. For lower values of luminosity at the surface, we are not able to find a solution. The temperature reaches a maximum and then exponentially drops on integrating inwards. This was seen in the envelope models of \cite{Stahler1988} (see Fig.~2 of that paper).
A way to think of this is that for high-luminosity envelopes, the envelope can accommodate a lower luminosity at the surface by reducing the base entropy, thereby reducing the luminosity entering the envelope at the base. However, at some point, the only way the envelope can accommodate a lower surface luminosity is by sending some of the compressional heating inwards through the lower boundary to the core. In Appendix~\ref{append B}, we describe an analytic model of the accreting envelope with a power law opacity that reproduces this behavior and helps to explain why accreting envelopes have a minimum luminosity.

\begin{figure}
\epsscale{1.15}
\plotone{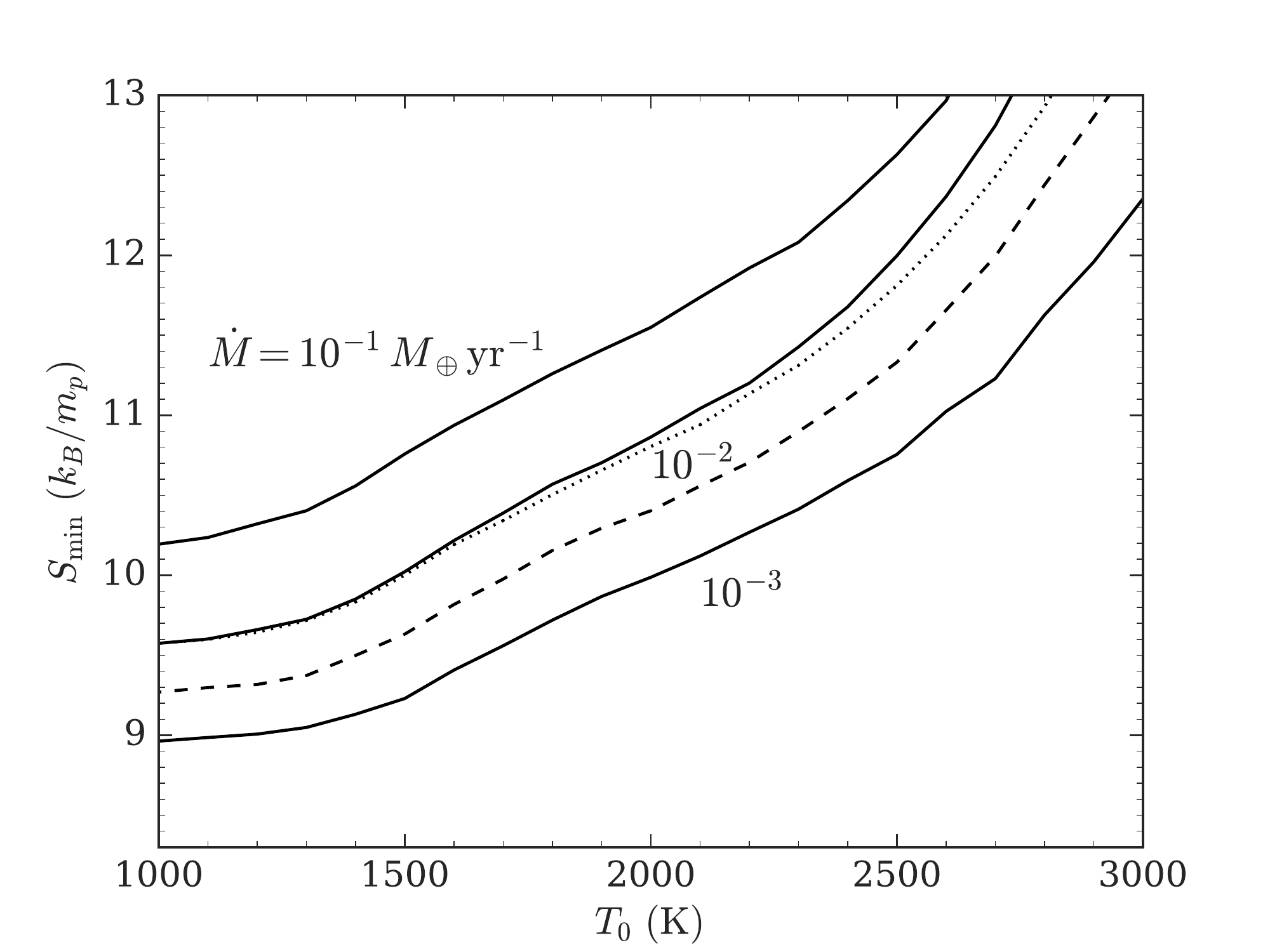}
\caption{The minimum value of internal entropy to which a radiative envelope can smoothly attach as a function of the outer boundary temperature $T_0$. The solid curves have the outer boundary pressure set equal to the ram pressure, with $M=1\ M_J$ and $R=2\ R_J$; the dotted curve shows the effect of increasing the outer boundary pressure by a factor of 10 for $\dot M=10^{-2}\ M_\oplus\,{\rm yr^{-1}}$. The dashed curve shows a more massive planet with $M=3\ M_J$, $R=1.5\ R_J$ accreting at $\dot M=10^{-2}\ M_\oplus\,{\rm yr^{-1}}$.}
\label{fig:Smin}
\end{figure}

To explore this further, we calculated the minimal entropy $S_{\rm min}$ at the base of the envelope as a function of $T_0$ and $\dot M$. Fixing $T_0$, we found the minimal-entropy envelope by solving for the luminosity at the surface that gave a vanishing luminosity at the base of the envelope. This solution is equivalent to the critical solution discussed by \cite{Stahler1988}; the minimal entropy $S_{\rm min}$ is equivalent to $s_{\rm sett}$ in that paper. Figure \ref{fig:Smin} shows how $S_{\rm min}$ varies with surface temperature $T_0$ for different accretion rates for a planet with mass $1\ M_J$ and radius $2\ R_J$. If the planet has an internal adiabat with $S_c>S_{\rm min}$, the radiative envelope can connect smoothly to the convective interior. This is not the case, however, if $S_c<S_{\rm min}$, implying that the accreted matter will accumulate with a much greater entropy than the internal adiabat. We explore the consequences of this in time-dependent models in \S~\ref{sec:mesamodels}. The value of $S_{\rm min}$ decreases with planet mass, which is shown by the dashed curve in Figure \ref{fig:Smin} which is for $\dot M=10^{-2}\ M_\oplus\ {\rm yr^{-1}}$ but for a $3\ M_J$, $1.5\ R_J$ planet. In calculating $S_{\rm min}$ we set the surface pressure to the ram pressure, but we find that $S_{\rm min}$ is not very sensitive to surface pressure (dotted curve in Fig.~\ref{fig:Smin}).

\begin{figure}
\epsscale{1.15}
\plotone{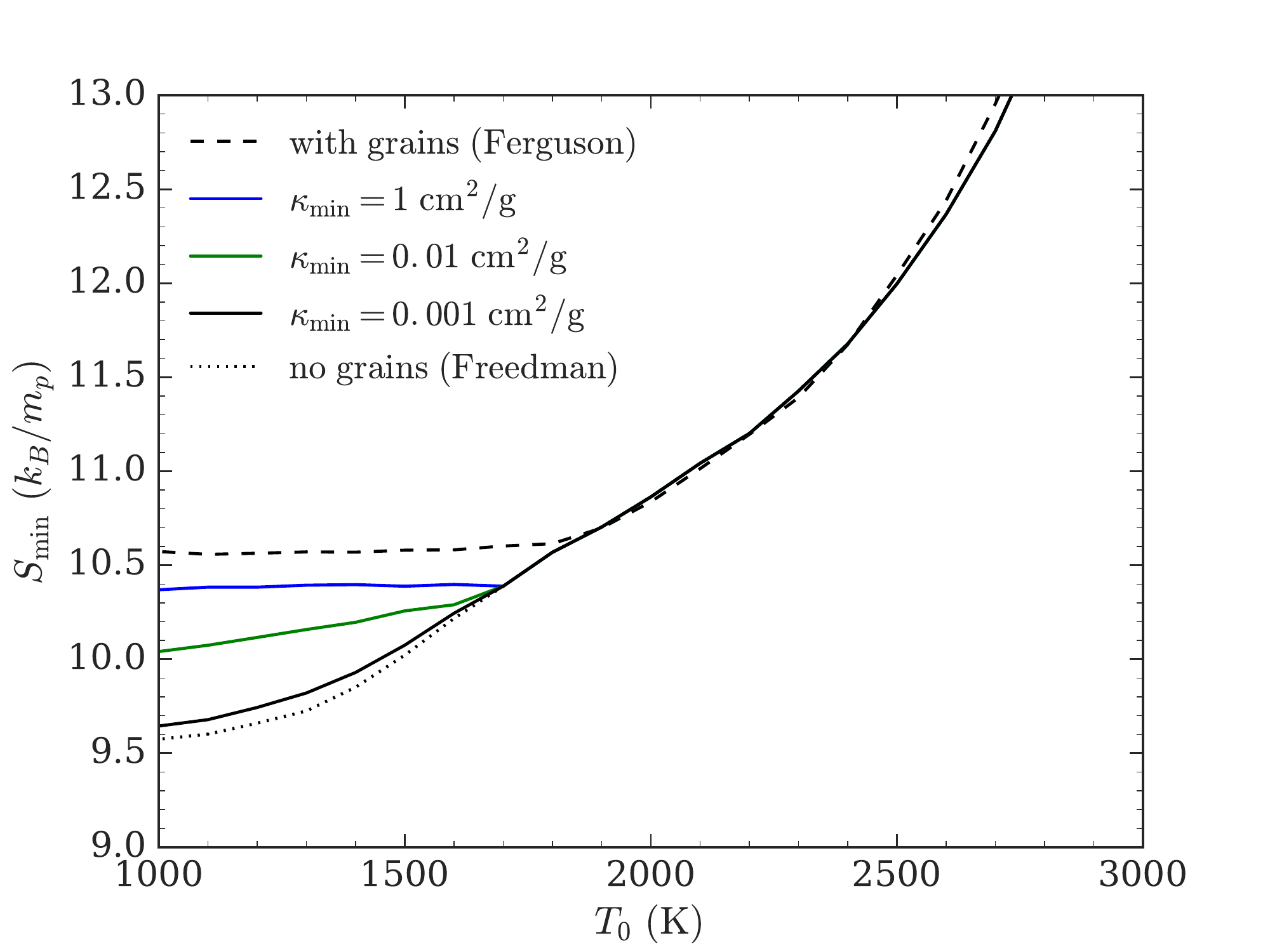}
\caption{The effect of grain opacity on the minimum entropy $S_{\rm min}$. The dotted curve is the $\dot M=0.01\ M_\oplus\ {\rm yr^{-1}}$ curve from Figure \ref{fig:Smin}. The other curves show the increase of $S_{\rm min}$ due to an increased opacity from grains at low temperatures.
}\label{fig:Smin2}
\end{figure}

\subsection{Influence of grain opacity}
\label{sec:grains}

To investigate the effect of grain opacity on the envelope, we use two approaches. First, to include the full grain opacity, we use the opacity tables from MESA based on the \cite{Ferguson2005} opacities (specifically the \texttt{lowT\_fa05\_gs98\_z2m2\_x70.data} table) for $X=0.7$ and $Z=0.02$. Second, we model a reduced grain opacity in an approximate way by adding a constant $\kappa_{\rm min}$ to the dust-free opacity tables from \cite{Freedman2008} for $T<1700\ {\rm K}$ (above approximately this temperature, grains evaporate, e.g.~\citealp{Semenov2003}). A reduction of grain opacity to about 0.3\% of the interstellar value \citep{Mordasini2014a} corresponds to $\kappa_{\rm min}\sim 10^{-2}\ {\rm cm^2\ g^{-1}}$.

We find that the additional opacity has two effects. The first is to increase the value of $S_{\rm min}$. This is shown in Figure \ref{fig:Smin2} for $\dot M=10^{-2}\ M_\oplus\ {\rm yr^{-1}}$. At temperatures below $1700\ {\rm K}$, the additional opacity in the envelope increases the value of $S_{\rm min}$ by $\approx 0.5\ k_{\rm B}/m_p$ for $\kappa_{\rm min}=10^{-2}\ {\rm cm^2\ g^{-1}}$. This shows that grain opacity can have an effect for accretion onto a planet with an initial value of internal entropy $S_i\lesssim 10.5\ k_{\rm B}/m_p$. In most of the cases we will show later, however, $S_{\rm min}$ becomes relevant only at higher temperatures where grain opacity is not important.

The second effect is that grain opacity acts to reduce the luminosity of cooling models. For example, a model with the same parameters as in Figure \ref{fig:profile} but $T_0=500\ {\rm K}$ (the lowest temperature available in the \citealt{Ferguson2005} tables) and $S_c=11.0\ k_{\rm B}/m_p$ has a luminosity at the RCB $L_{\rm RCB}=6.9\times 10^{-4}\ L_\odot$ with molecular opacity only (opacities of \citealt{Freedman2008}) and $L_{\rm RCB}=6.9\times 10^{-5}\ L_\odot$ with full grain opacity (opacities of \citealt{Ferguson2005}). Setting $\kappa_{\rm min}=10^{-2}\ {\rm cm^2\ g^{-1}}$ gives $L_{\rm RCB}=2.1\times 10^{-4}\ L_\odot$, a few times lower than the grain-free case. 

Both of these effects make it harder to produce cold starts. For the rest of the paper we use only the grain-free molecular opacities of \cite{Freedman2008}, keeping in mind that in the cooling regime dust opacity will act to increase the final entropy, and so we are being optimistic for the production of cold starts.

\begin{figure}
\epsscale{1.2}
\plotone{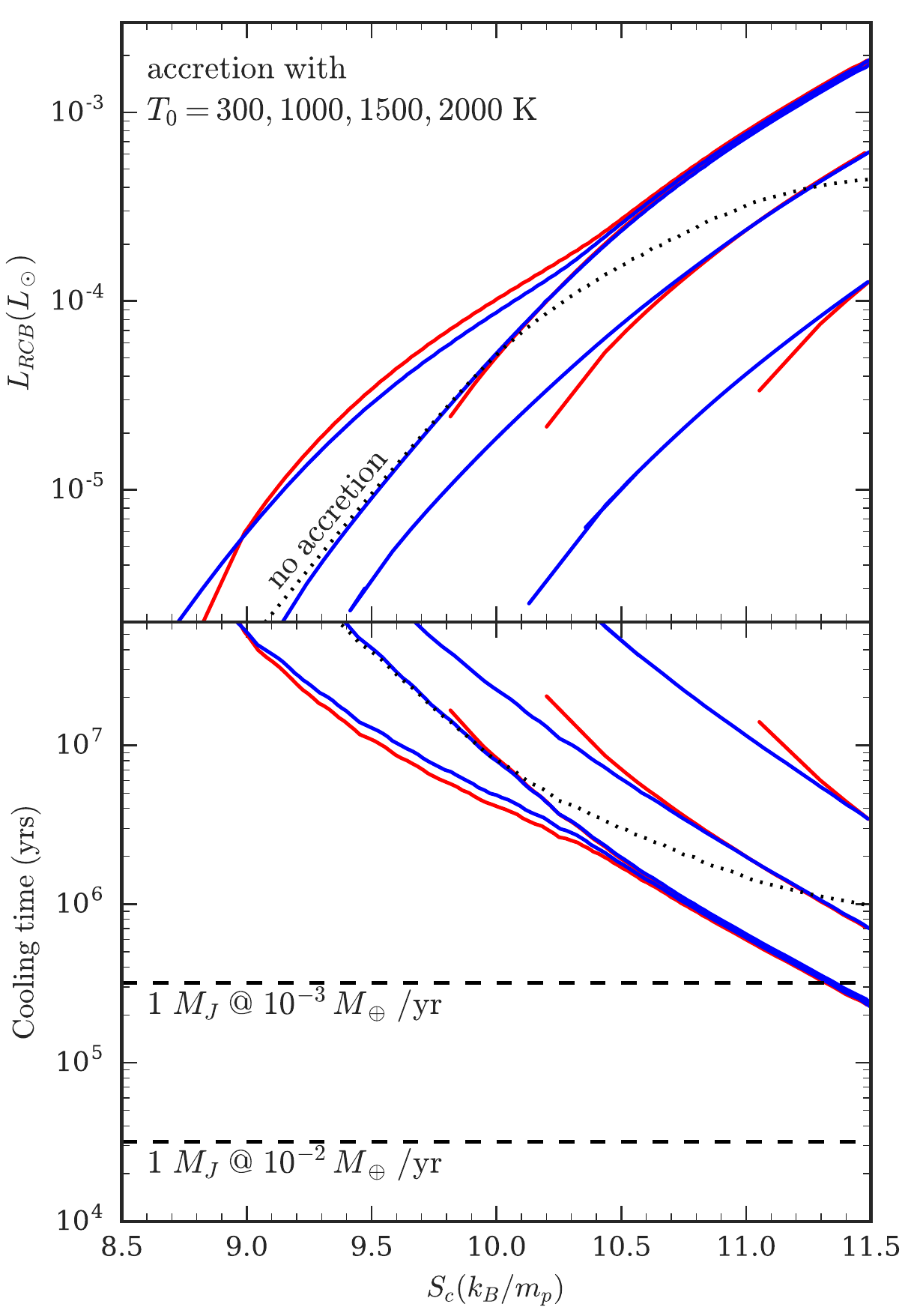}
\caption{Luminosity at the radiative-convective boundary (\textit{upper panel}) and cooling timescale (\textit{lower panel}) as a function of the entropy of the convection zone for accreting models with outer temperatures $T_0=300$, 1000, 1500, and 2000~K (\textit{from left to right}) and $\dot M=10^{-2}\ M_\oplus\ {\rm yr^{-1}}$ (\textit{red curves}) and $10^{-3}\ M_\oplus\ {\rm yr^{-1}}$ (\textit{blue curves}). The dotted curve shows the luminosity and cooling time of an isolated (non-accreting) planet. The planet mass is $1\ M_J$, and for each value of $S_c$ we set the appropriate radius and the outer boundary pressure to the ram pressure (eq.~[\ref{eq:Paccr}]). The horizontal dashed lines in the lower panel show the time to accrete $1\ M_J$ for each accretion rate.
}
\label{fig:LS}
\end{figure}

\subsection{Cooling timescales during accretion}
\label{sec:cooling timescales}

Figure \ref{fig:LS} shows the cooling luminosity $L_{\rm RCB}$ for a range of model parameters. The different curves show $L_{\rm RCB}$ as a function of the internal entropy $S_c$ for different boundary temperatures. 
The dotted curve shows the luminosity of an isolated planet for comparison. The accreting models with $T_0=1000\ {\rm K}$ and smaller are more luminous than the isolated planets; those with $T_0=1500$ or $2000\ {\rm K}$ are less luminous than an isolated planet.

The models shown are for a specific choice of planet mass, $M=1\ M_J$, but can easily be rescaled to other masses. For radiative envelopes,  $L$ and $M$ enter the radiative diffusion equation in the combination $L/M$ (see eq.~[\ref{eq:raddiffusion}]; \citealt{Arras2006,Marleau2014}) and so $L_{\rm RCB}\propto M$. We find this scaling is a good approximation for all our models.

Two different accretion rates are shown in Figure \ref{fig:LS}. The cooling luminosity $L_{\rm RCB}$ does not depend sensitively on $\dot M$. The main effect of changing the accretion rate is to change the minimum value of entropy $S_{\rm min}$ for which we can have a cooling core. Each curve in Figure \ref{fig:LS} starts at $S_{\rm min}$ (compare Fig.~\ref{fig:Smin}). For example, the models with $T_0=1000\ {\rm K}$ only allow a cooling envelope attached to the interior convection zone for $S_c\gtrsim 9.8\ k_{\rm B}/m_p$ at $\dot M>0.01\,M_\oplus\ {\rm yr^{-1}}$. At smaller internal entropies, the planet will accumulate a hot envelope with entropy $\approx 10\ k_{\rm B}/m_p$, which is a potentially much higher entropy than in the convective core.

The lower panel of Figure \ref{fig:LS} shows the cooling time of the planet. We calculate the cooling time by taking the cooling time of an isolated gas giant with the same mass and entropy $S_c$ from \cite{Marleau2014} and scaling it by the ratio of the RCB luminosity in the accreting envelope to the RCB luminosity without accretion. An outer temperature of 300~K reduces the cooling time by a factor of a few or more; hotter boundaries with $T_0\gtrsim 1500\ {\rm K}$ have longer cooling times than an isolated planet by factors of a few to an order of magnitude.

To reduce the internal entropy substantially during accretion, and therefore make a cold start, the cooling timescale should be shorter than the accretion time. It is striking in Figure \ref{fig:LS} that this is almost never the case. Even for a cold outer boundary $\lesssim 1000\ {\rm K}$ and accretion rate $10^{-3}\ M_\oplus\ {\rm yr^{-1}}$, the cooling time is still comparable to the accretion time. To give a specific example, for an accretion rate of $10^{-2}\ M_\oplus\ {\rm yr^{-1}}$ the accretion time is $\approx 3\times 10^4\ {\rm yr}$ per Jupiter mass. For internal entropy $S_c\gtrsim 11\ k_{\rm B}/m_p$, this is a factor of $\gtrsim 3$ times longer than the cooling timescale, so while some cooling can occur we would not expect a large change in entropy during accretion. At $10^{-3}\ M_\oplus\ {\rm yr^{-1}}$, an entropy $11.5\ k_{\rm B}/m_p$ object has a cooling time shorter than the accretion time for $1\ M_J$ and so should be able to cool as it forms, but we would not expect it to be very dramatic. For hotter boundaries with $T_0\gtrsim 1500\ {\rm K}$, the cooling is effectively stalled by the hot envelope.


\section{Evolution of the planet during accretion with MESA}
\label{sec:mesamodels}

In section \ref{sec:theorymodels} we studied time-independent snapshots of planetary models. We now use the open-source 1D stellar evolution code MESA\footnote{Modules for Experiments in Stellar Astrophysics, version 7623} \citep{Paxton2011,Paxton2013,Paxton2015} to model the time-dependent evolution of a planet during runaway gas accretion. The implementation and evolution of the model\footnote{Input files for our set-up can be found at \url{http://mesastar.org}.} is given in \S~\ref{sec:mesa setup}.  We first adopt a constant temperature $T_0$ and pressure $P_0$ at the surface of the planet (\S~\ref{sec:acc regimes}) to explore the influence of the entropy $S_0=S(T_0,P_0)$ of the accreted material on the final state of the planet. We then adopt a more realistic time-dependent outer boundary condition where we set the pressure to the ram pressure and parameterize the outer boundary in terms of the shock temperature $T_0$ (\S~\ref{sec:ram pressure}).

\subsection{Details of Planet Model and Simulating Accretion}
\label{sec:mesa setup}

\subsubsection{Starting model}
\label{sec:initial planet models}

We create an initial planet model for accretion using the \texttt{make\_planet} test suite in MESA. We set the mass and radius of the planet, leaving other parameters at their default values but turning off irradiation. The hydrogen and helium mass fractions are $X = 0.73$, and $Y = 0.25$ respectively, the low-temperature opacity tables are those of \cite{Freedman2008}, and the equation of state is given by \cite{Saumon1995}. We include a rocky core with mass and radius $10\ M_\oplus$ and $2.8\ R_\oplus$ (i.e.~with a mean density of $10\ {\rm g\ cm^{-3}}$), which is implemented in MESA through simple inner boundary conditions for the structure of the modeled planet.

For a given initial mass of the planet, we choose the radius in order to set the desired initial internal entropy. In the core accretion models of \cite{Mordasini2013}, the entropy of the planet at the onset of runaway accretion is $\approx 11\ k_{\rm B}/m_p$. To explore the sensitivity to the starting entropy, we consider values of $S_i=9.5$, $10.45$ and $11.6\ k_{\rm B}/m_p$. At these values of entropy, the \texttt{make\_planet} module has difficulty converging for masses as low as the crossover mass $\lesssim 0.1\ M_J$ because the planet is greatly inflated. To alleviate this problem, we instead start with a larger mass, $0.2$, $0.5$, and $1\ M_J$ for $S_i=9.5$, $10.45$ and $11.6\ k_{\rm B}/m_p$, respectively\footnote{We investigated the sensitivity to changing the initial mass, and found that the final entropy of the planet changed by $\lesssim 0.3\ k_{\rm B}/m_p$.}. For these three choices of initial mass, we set the radius in \texttt{make\_planet} to $R=2$, $5$, and $10\ R_J$, which leads to the desired entropy at the onset of accretion.

\subsubsection{Accretion and the Outer Boundary Conditions}
\label{sec:shock boundary conditions for accretion}

We now turn on accretion using the \texttt{mass\_change} control to specify an accretion rate. By default, MESA accretes material with the same thermodynamic properties (i.e.~temperature, density and thus entropy) as the outer layers of the model. This is a useful comparison case which we will refer to as thermalized accretion. To model runaway gas accretion, we use the \texttt{other\_atm} module of the \texttt{run\_star\_extras} file in MESA in order to specify $T_0$ and $P_0$.
They can be set for example to constant values for the entire evolution, or adjusted depending on the state of the planet at any given time (e.g.~the mass- and radius-dependent ram pressure given by eq.~[\ref{eq:Paccr}]).

If the deviation from thermalized accretion is too large, MESA may fail to converge and not produce a model. Consequently, if the imposed surface temperature is too high, we slowly increase the temperature from a lower value that does converge to the desired temperature over a timescale on the order of $\sim  1\%$ of the total accretion time to ensure that the final results are not significantly affected. 
For example, a model accreting at a rate of $10^{-2}\ M_\oplus\ {\rm yr}^{-1}$ with a desired surface temperature of 2500~K will instead begin with 1500~K and linearly increase the temperature up to 2500~K over the course of 5000~yr. 

We do not include any internal heating from planetesimal accretion. Planetesimals can deposit energy deep inside the planet, with maximal luminosity when they penetrate to the rocky core (e.g.~see discussion in  \S~5.7 of \citealt{Mordasini2015}). The luminosity is $L_Z=(GM_c/R_c)\dot M_Z\approx 10^{-6}\ L_\odot\ (\dot M_Z/10^{-5}\ M_\Earth\ {\rm yr^{-1}})$, where $\dot M_Z$ is the accretion rate of planetesimals and we take a core mass $M_c=10\ M_\Earth$ and mean core density $\bar\rho_c=5\ {\rm g\ cm^{-3}}$. Because it is deposited potentially deep inside the convection zone, this luminosity can heat the convection zone from below and cause its entropy to increase. However, the internal luminosities we find are all much greater than $L_Z$, except for the coldest cases, and so we neglect this heat source.

As a check that the MESA calculations are converging to a physical model, we increase and decrease by a factor of two the \texttt{mesh\_delta\_coeff} parameter, which controls the length of the grid cells, and find no discernible difference in the results. Similarly, we lower by an order of magnitude the \texttt{varcontrol\_target} parameter, which controls the size of the time step, and again find no difference.

\subsection{Identification of accretion regimes}
\label{sec:acc regimes}

We first survey the final entropies obtained by holding $T_0$ and $P_0$ fixed during accretion. We construct a grid of models with $T_0$ and $P_0$ ranging from $100$ to $2700$~K and $10^{2.3}$ to $10^{5.5}$~$\rm{erg\ cm^{-3}}$ respectively. For these values the surface entropy $S_0$ ranges from $\approx6$ to $20\ k_{\rm B}/m_p$ (Appendix~\ref{append A}). In this section, we use an accretion rate of $10^{-2}\ M_\earth\ {\rm yr}^{-1}$, an initial mass of 0.5 $M_J$, and an entropy of 10.45 $k_{\rm B}/m_p$.

\begin{figure}
\epsscale{1.2}
\plotone{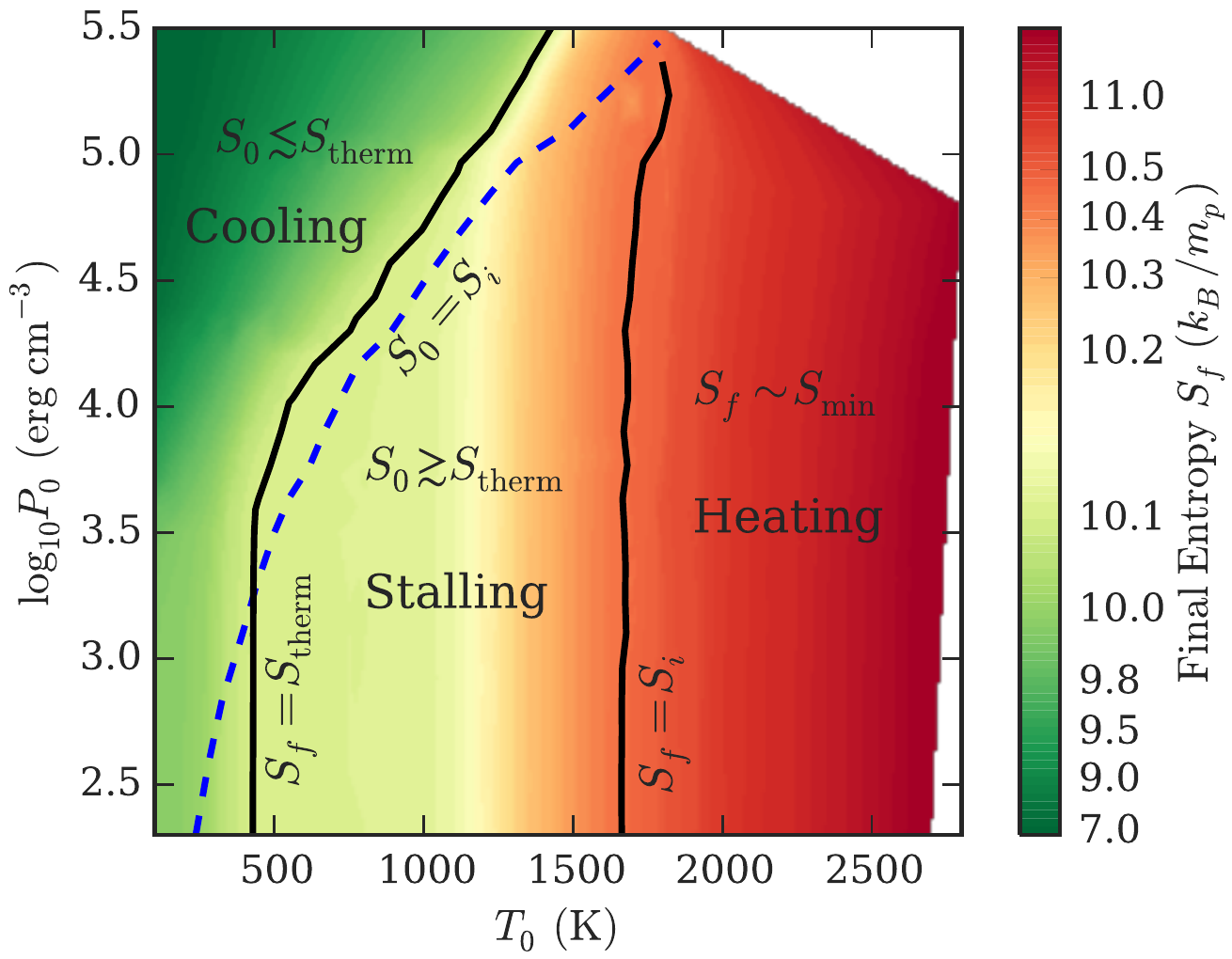}
\caption{Final entropy (\textit{colorscale}) of a $10\ M_J$ planet accreting at $10^{-2}\ M_\oplus\ \rm{yr}^{-1}$ as a function of surface temperature $T_0$ and pressure $P_0$, held constant. Every model begins with a mass of $0.5\ M_J$ and an initial entropy of $S_i = 10.4\ k_{\rm B}/m_p$. The black line on the right indicates where the final entropy $S_f$ is equal to $S_i$. The black line on the left indicates where the final entropy is equal to the entropy reached by thermalized accretion $S_{\mathrm{therm}} = 10.1\ k_{\rm B}/m_p$. The blue dashed line indicates where the surface entropy $S_0$ is equal to the initial entropy. The three accretion regimes (``cooling'', ``stalling'', and ``heating'') are discussed in the text. The colors and contours were obtained by smoothing an appropriately-distributed set of 989 independent models.}
\label{fig:TP grid scan}
\end{figure}

The results of this survey are shown in Figure \ref{fig:TP grid scan}.
We find that the final entropies can be separated into three different regimes. The black line on the right shows where the final entropy of the planet at the end of accretion is equal to the initial entropy. In the region to the right of this line the final entropy is greater than the initial entropy, hence the `heating' regime. In the region to the left of this line, the final entropy is lower than the initial entropy, and this can be further subdivided into two more regions.

The black line in the left of Figure \ref{fig:TP grid scan} shows where the final entropy of the planet is equal to the value it would reach under thermalized accretion, in which the accreted material has the same thermodynamic properties as the planet. In a sense, this scenario allows the planet to cool while increasing its mass. The final entropy reached under this conditions is referred to as $S_\mathrm{therm}$. It can be seen that in most cases, if $S_0 > S_\mathrm{therm}$ then the final entropy of the planet will be between $S_i$ and $S_\mathrm{therm}$, in the `stalling' regime, since the planet has not cooled as much as it could have. To the left of the leftmost black line, we have the region where $S_f < S_\mathrm{therm}$, which is again characterized by having $S_i < S_\mathrm{therm}$. In this `cooling' regime, the planet cools by a greater amount than it would have and thus ends up at a lower final entropy.

\begin{figure}
\epsscale{1.15}
\plotone{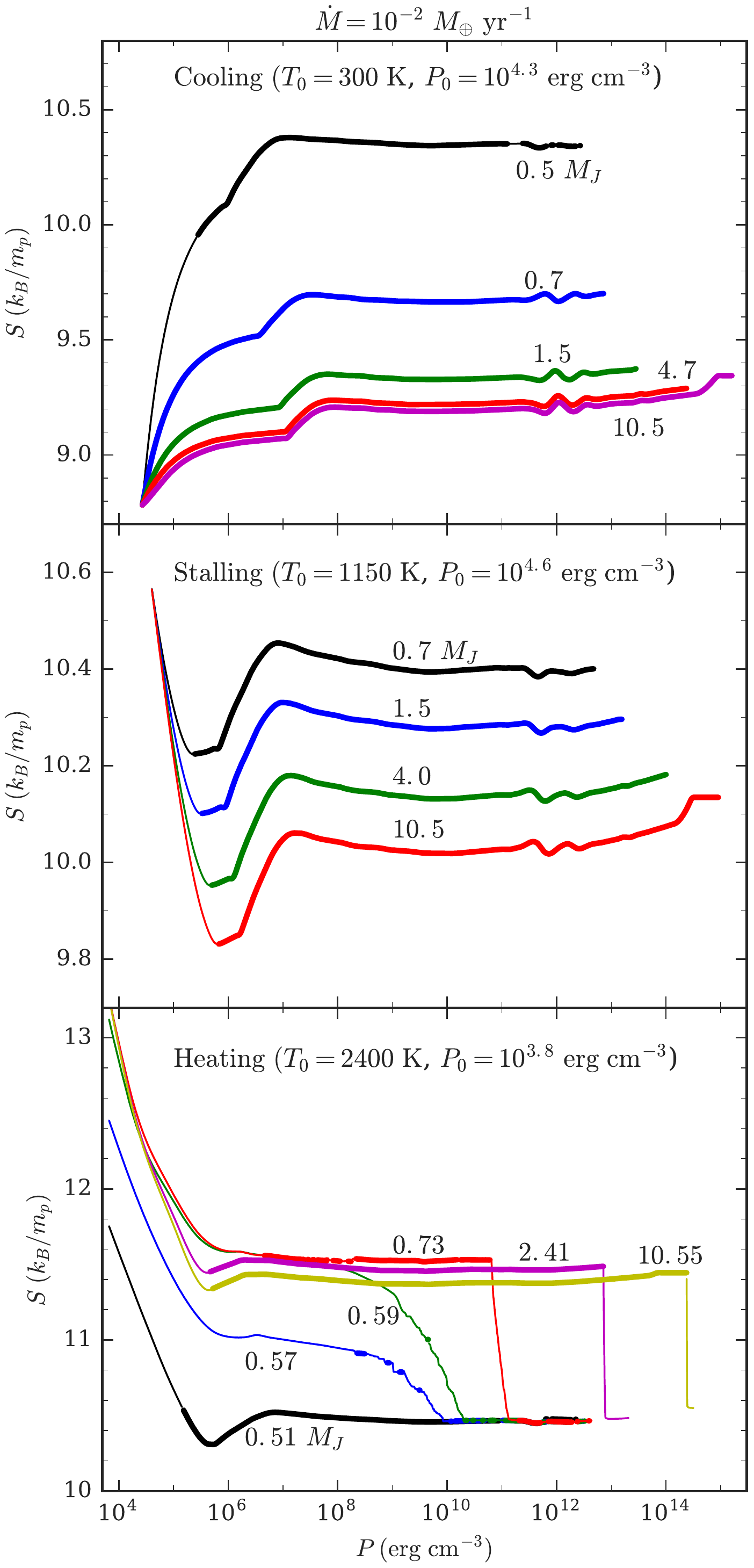}
\caption{Internal entropy profiles for a planet with initial entropy $S_i =10.45\ k_{\rm B}/m_p$ undergoing accretion with boundary conditions ($T_0$ and $P_0$). They are chosen to correspond to the three accretion regimes identified in Figure~\ref{fig:TP grid scan} (\textit{see panel titles}), with entropies for the accreted material of respectively $S_0 = 8.7$, 10.6, and~$13\ k_{\rm B}/{m_p}$ (\textit{top to bottom panel}). The total mass (\textit{labels next to curves}) is used to track the time evolution of the models from 0.5 to 10.5~$M_J$. Convective regions in the profiles, according to the Schwarzschild criterion, are shown by thick lines. Note that each panel uses a different scale on the vertical axis.}
\label{fig:entropy_profiles}
\end{figure}

  In Figure \ref{fig:entropy_profiles}, we look at the internal profiles for planets accreting in each regime at different points throughout their accretion, in order to understand what drives their evolution. The top panel shows the evolution under `cooling' accretion conditions, where the surface entropy is at a value of $S_0 \approx 8.7\ k_{\rm B}/m_p$, which is below $S_\mathrm{therm} = 10.1\ k_{\rm B}/m_p$. We see the internal entropy decreases rapidly, such that it drops to almost the surface entropy $S_0$ after accretion of about one Jupiter mass or about 30,000 years. This corresponds to the cold-outer-boundary envelope discussed in \S~\ref{sec:cooling boundary model}. The internal entropy structure is such that the entire planet is convective as it cools down.

The middle panel of Figure \ref{fig:entropy_profiles} shows the {\em stalling} regime, in which the surface entropy is higher than $S_\mathrm{therm}$, but still low enough to smoothly attach to the interior of the model (\S~\ref{sec:envelope models}). A radiative region forms in the outer layers, which pushes the RCB to higher pressures, reducing the luminosity from the convective core (\S~\ref{sec:cooling timescales}). The internal entropy still decreases, but at a slower rate than in the cooling scenario or thermalized accretion.

The bottom panel of Figure \ref{fig:entropy_profiles} shows the {\em heating regime}, in which the difference in entropy between the surface and interior is too large for the envelope to accommodate, as discussed in \S~\ref{sec:hot boundary model}. In this case, the accreted material accumulates to form a second convection zone above the original convective core. Note that there is a temperature inversion associated with the jump between the original low convective entropy zone and the new, higher-entropy convection zone; a similar temperature inversion was seen for strongly-irradiated hot jupiters by \cite{Wu2013}. The conduction timescale in the planet interior is very long, so that the temperature inversion remains at the same mass coordinate as accretion proceeds. As mentioned in \S~\ref{sec:shock boundary conditions for accretion}, the surface temperature is increased linearly from 1500~K to 2400~K over the course of 5000 years to help convergence. This gives the initial rise of the surface entropy for $M\lesssim 0.7\ M_J$.

\begin{figure}
\epsscale{1.15}
\plotone{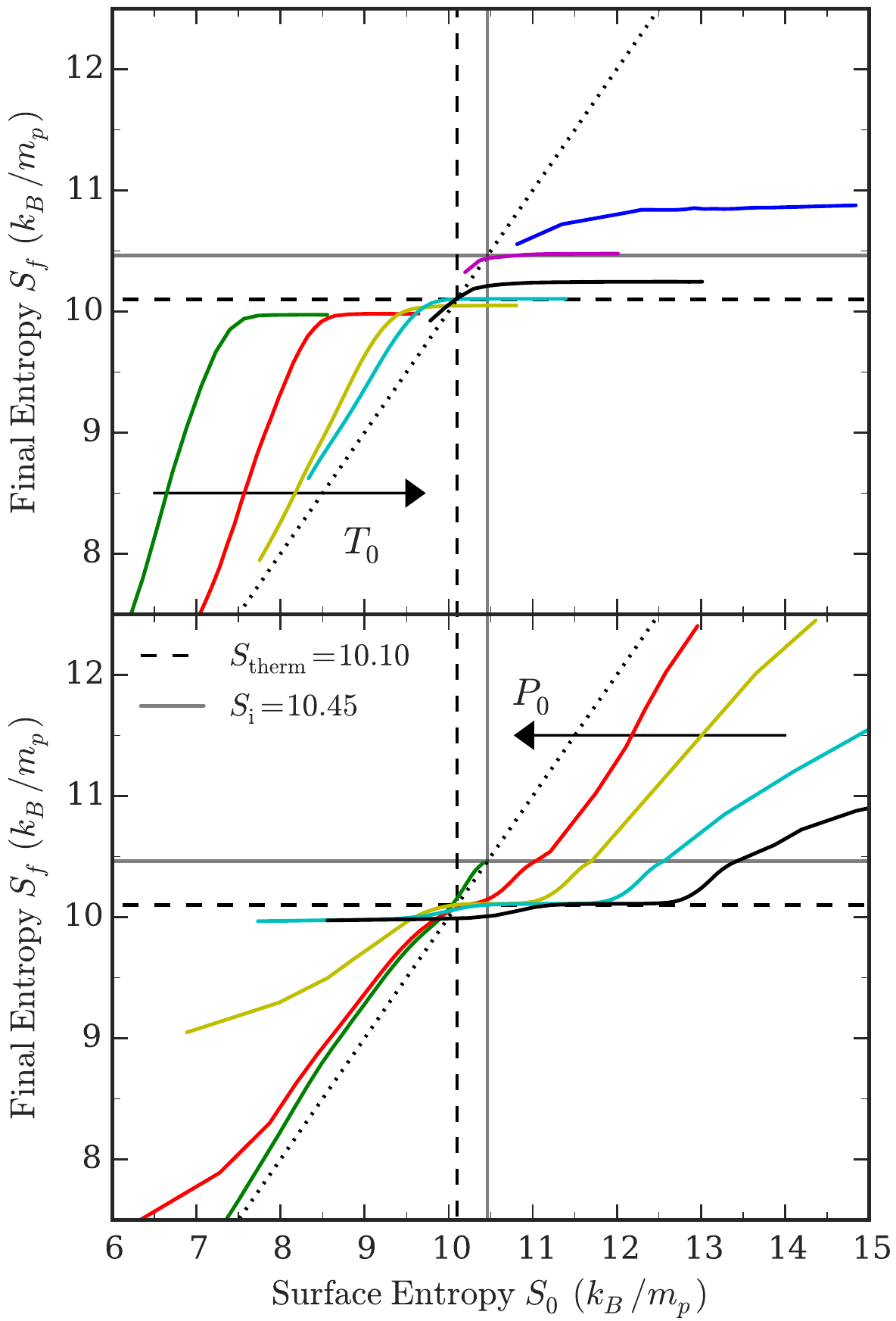}
\caption{\textit{Top panel:} Final internal entropy of the planet as a function of the entropy of the accreted surface material. The models are as in Figures~\ref{fig:TP grid scan} and~\ref{fig:entropy_profiles}. For structures with two convective zones, the entropy of the upper zone is used, as discussed in the text. The colored lines correspond to constant values of the shock temperature $T_0 =100$, 150, 300, 450, 1350, 1750, 2100~K (\textit{bottom left to top right}). Along each constant-$T_0$ curve, the surface pressure $P_0$ decreases from left to right. Displayed are also the value of the initial entropy of the model ($S_i=10.46\ {k_{\rm B}/m_p}$; \textit{solid gray line}) and the final entropy reached with thermalized accretion ($S_{\rm therm} = 10.10\ {k_{\rm B}/m_p}$; \textit{dashed black line}). The diagonal dotted line shows where the final and surface entropy are equal. \textit{Bottom panel:} Same results as in the top panel but plotted as curves of constant shock pressure $P_0$ for $\log_{10}(P_0/\mathrm{erg\ cm^{-3}}) = 2.3$, 3.2, 4.1, 4.8, 5.5 (\textit{top right to bottom left}); along each curve, the shock temperature $T_0$ increases from left to right.}
\label{fig:center_v_surf}
\end{figure}

To see how the boundary conditions determine the post-accretion planet properties, Figure~\ref{fig:center_v_surf} shows the final interior entropy $S_f$ as a function of the surface entropy $S_0$ for a final planet mass of $10\ M_J$. In the hot models that develop two internal convection zones, we choose the higher internal entropy value since most of the mass of the planet is at this higher entropy value. This in turn is due to the upper zone appearing sufficiently early in the accretion history; for instance, in Fig.~\ref{fig:entropy_profiles}, only the inner $\approx 0.5\ M_J$ are frozen in at $S\approx S_i=10.45\ k_{\rm B}/m_p$.

Models with $S_0<S_\mathrm{therm}$ (to the left of the dashed vertical line in Figure~\ref{fig:center_v_surf}) are in the cooling regime. They show that the amount of cooling at a given value of surface entropy $S_0$ depends on the explicit choice of $P_0$ and $T_0$. Also, in this regime there is a stronger dependence on pressure than on temperature. 
For a fixed surface entropy, moving the surface to higher pressure means that the entropy must increase at a faster rate to match onto the internal value, implying a larger value of $\nabla-\nabla_{\rm ad}\propto dS/dP$ and therefore a larger convective luminosity (eq.~[\ref{eq:Lconv}]). A higher surface pressure therefore gives more rapid cooling, resulting in a lower value of $S_f$ at the end of accretion. It should be noted that cooling below $9\ k_{\rm B}/{m_p}$ requires high pressures ($P_0>10^{4.2}\ \mathrm{erg\ cm^{-3}}$) and low temperatures ($T_0< 450\ \mathrm{K}$).

For $S_0>S_\mathrm{therm}$, we see the stalling and heating regimes. In the heating regime, the final entropy lies above the initial entropy, and increases with $T_0$, having almost no dependence on $P_0$. In the stalling regime, the final entropy lies between the initial value $S_i$ and $S_\mathrm{therm}$. 
As $T_0$ increases in the stalling regime, the RCB is pushed to higher pressure, reducing the luminosity at the RCB and delaying the cooling further so that the final entropy of the planet is approximately equal to the initial entropy $S_i$. This is a similar effect to the delayed cooling of irradiated or Ohmically-heated hot jupiters (e.g.~\citealt{Arras2006,Huang2012,Wu2013}). In this regime, the degree of cooling is insensitive to $P_0$ because the envelope is close to isothermal (e.g.~see Fig.~\ref{fig:profile}), so that it is the temperature of the envelope set by $T_0$ that determines the RCB location. 

Additionally, the same grid of $T_0$ and $P_0$ was run for an initial entropy $S_i=11.5\ k_{\rm B}/m_p$. The final entropy reached under thermalized accretion was essentially the same, since for high initial entropies this value will be set by the amount of time available to cool. Since the heating/stalling boundary is located at the initial entropy, this only increased the `height' of the stalling regime, i.e.\ the distance between the horizontal lines in Figure \ref{fig:center_v_surf}.

\subsection{The outcome of runaway accretion}

\label{sec:ram pressure}

In order to model runaway accretion, we now use the ram pressure $P_{\rm accr}$, given by equation (\ref{eq:Paccr}), as the outer boundary pressure $P_0$. The ram pressure evolves with time as the mass and radius of the planet change. We hold the outer temperature $T_0$ constant. In reality, the shock temperature will also depend on mass and radius and change with time
(e.g.~as in eq.~[\ref{eq:hotT}]), but without a specific model, we leave it as a constant parameter describing the post-shock conditions (\S~\ref{sec:summdisc}). 

\begin{figure}
\epsscale{1.3}
\plotone{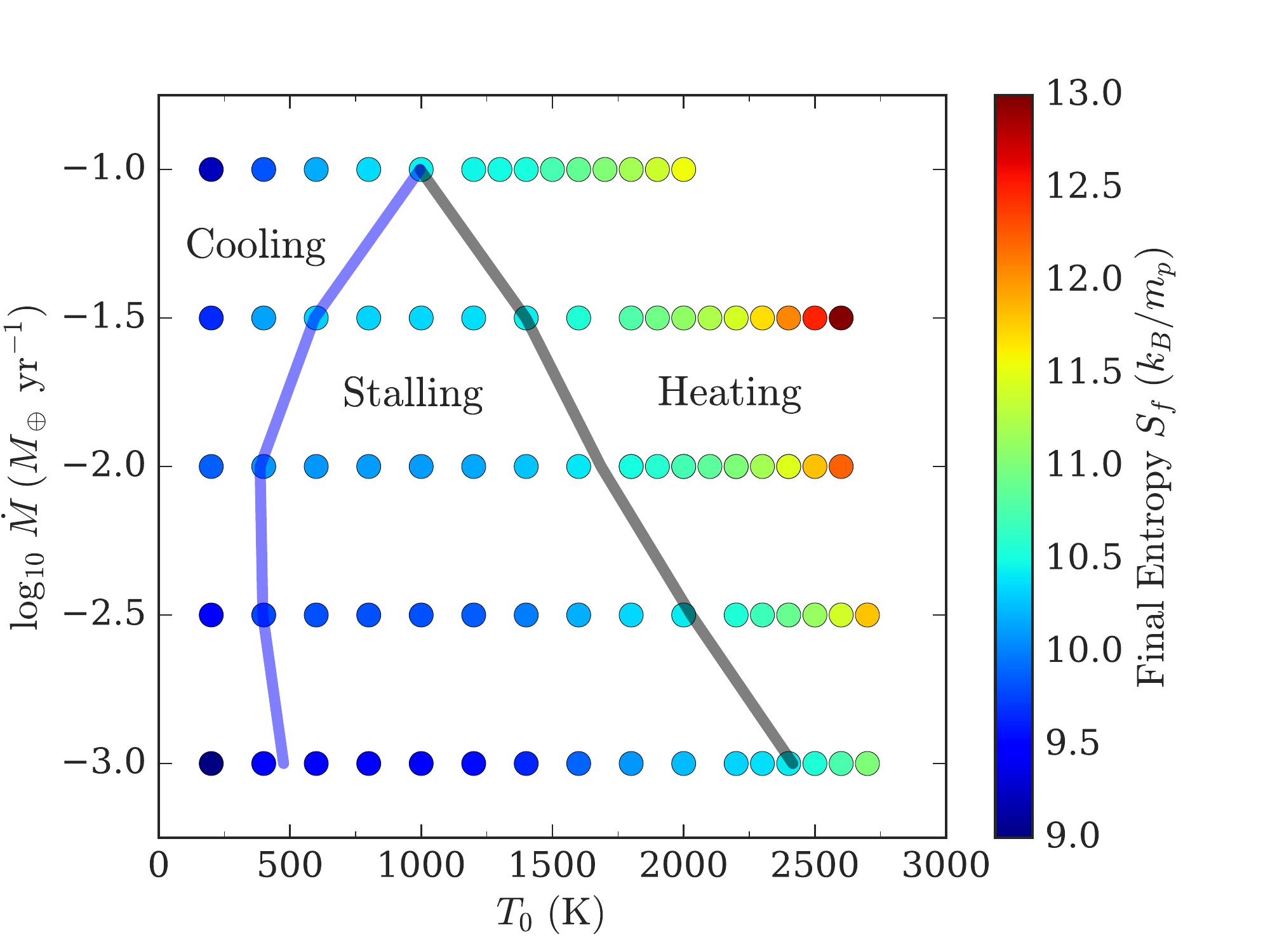}
\caption{Final internal entropy (\textit{colorscale}) of the planet as a function of shock temperature $T_0$ and accretion rate $\dot{M}$. The solid black line indicates the initial entropy of the models (here $S_i = 10.45\ k_{\rm B}/m_p$), thus delineating the stalling and heating regimes. The solid blue line indicates the final internal entropy reached under thermalized accretion, separating the cooling and stalling regimes. This value depends on the accretion rate, so that along the blue line the entropy value changes.}
\label{fig: ram entropy results} 
\end{figure}

Figure \ref{fig: ram entropy results} shows the final central entropy of the planet as a function of $T_0$ and $\dot{M}$, having started with entropy $S_i= 10.45\ k_{\rm B}/m_p$. We again see the separation into three accretion regimes. The blue line is drawn such that the entropy along it is at the value that would be reached by thermalized accretion at each accretion rate. The entropies to the left of the blue line are smaller, indicating the cooling regime. The black line is drawn such that the entropy along it is equal to the initial entropy. The entropy to the right of the black line are greater, indicating the heating regime. Between the blue and black lines, where the entropy lies between the initial value and the value reached by thermalized accretion, is the stalling regime.

In the cooling regime, the entropy reaches a minimum of $\sim$ 9 ${k_{\rm B}/m_p}$, whereas we found much lower values in \S~\ref{sec:acc regimes}. The difference is due to the fact that the ram pressure never gets high enough to decrease the surface entropy significantly. For example with $\dot{M} = 10^{-2}\ M_\oplus\ {\rm yr^{-1}}$ and a
final radius $R \approx 1\ R_J$ and mass $M=10\ M_J$, the ram pressure is always $P_{\rm accr} \lesssim 10^4\ {\rm erg\ cm^{-3}}$ since $P_{\rm accr}\propto M^{1/2}R^{-5/2}$ (eq.~[\ref{eq:Paccr}]);
comparing to Figure \ref{fig:center_v_surf}, this does not lead to significant cooling.

The internal entropy in the cooling regime depends in a non-monotonic way on the accretion rate. Increasing the accretion rate from $10^{-2}$ to $10^{-1}\ M_\oplus\ {\rm yr^{-1}}$ yields a lower entropy because the ram pressure is higher for a higher accretion rate, leading to a larger luminosity (Fig.~\ref{fig:center_v_surf}). At lower accretion rates $\dot M\gtrsim 10^{-3}\ M_\oplus\ {\rm yr^{-1}}$, the luminosity is smaller than at $\dot M\gtrsim 10^{-2}\ M_\oplus\ {\rm yr^{-1}}$, but the accretion timescale is much longer so that more cooling can occur and the final entropy decreases with decreasing $\dot M$. For $\dot M\gtrsim 10^{-2}\ M_\Earth\ {\rm yr^{-1}}$, the boundary between the cooling and stalling regimes is at larger temperature for larger accretion rate. This is because the ram pressure is larger, and a higher temperature is needed to have a large enough entropy to be in the stalling regime. For $\dot M\lesssim 10^{-2}\ M_\Earth\ {\rm yr^{-1}}$, the boundary temperature is almost independent of accretion rate, because the boundary moves to low pressure (horizontal parts of the curves in the top panel of Fig.~\ref{fig:center_v_surf}).

In the stalling regime, the final entropy increases with accretion rate because there is less time available to cool, and increases with temperature because  a hotter envelope reduces the cooling luminosity. In the heating regime, the final entropy is set by $S_{\rm min}$, which increases with temperature and accretion rate. The values of entropy agree well with the values of $S_{\rm min}$ calculated in the envelope models (Fig.~\ref{fig:Smin}). The boundary between the stalling and heating regimes can be understood by finding the temperature for which $S_{\rm min}\approx S_i$ at each $\dot M$.

\begin{figure*}
\epsscale{1.2}
\plotone{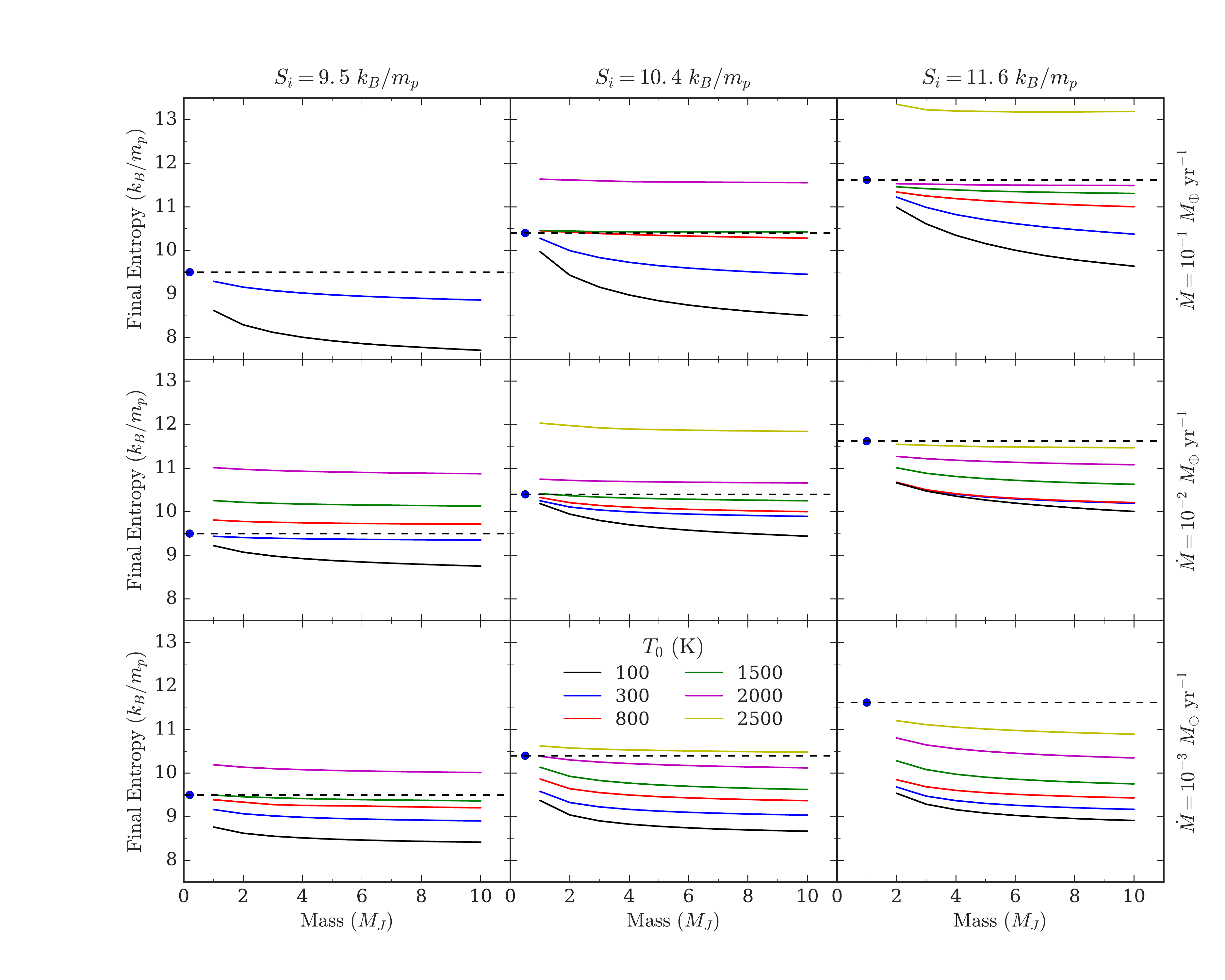}
\caption{Final entropy as a function of mass for accretion models. Each panel shows a particular choice of $\dot{M}$ and $S_i$ indicated by the labels along the top and right of the figure. The blue dots and dashed lines indicate the initial entropy and mass, which are (9.5, 0.2), (10.4, 0.5), and (11.6, 1.0) (${k_{\rm B}/m_p}$, $M_J$) from the left column to the right column. The lines correspond to accretion with different surface temperature $T_0$ (\textit{see legend}). Not all temperatures are shown in some panels because of convergence issues at lower values of $S_i$ and larger values of $\dot M$ or $T_0$.}
\label{fig: ram pressure tuning forks}
\end{figure*}

Figure \ref{fig: ram pressure tuning forks} shows, for different values of $S_i$, $\dot M$, and $T_0$, the dependence of the internal entropy on planet mass, i.e.~the post-formation, initial entropy (`initial' in terms of the pure cooling phase; e.g.~\citealp{Marley2007}). In each panel, the blue dot shows the initial mass and entropy. For the cooling cases, the curves drop rapidly with increasing mass at first but then flatten at larger masses. Most of the cooling happens by the time that they have reached $\approx 4\ M_J$ (as can also be seen in the entropy profiles in Fig.~\ref{fig:entropy_profiles}). The models in the heating regime show a final entropy that depends only slightly  on total mass ($\Delta S\approx0.2\ k_{\rm B}/m_p$ from~1 to 10~$M_J$ at a given $T_0$). In these cases, immediately after accretion starts the hot envelope deposits matter with entropy $S_{\rm min}$ in a second convection zone as described in the \S~\ref{sec:acc regimes}. However, Figure \ref{fig:Smin} shows that $S_{\rm min}$ decreases with planet mass, so that very quickly the planet enters the stalling regime where the accreting envelope joins smoothly onto the high-entropy outer convection zone. This lets internal entropy decrease slightly with planet mass after the initial rise. This result differs from the hot-start accretion models of \cite{Mordasini2013}, which show an increasing entropy with mass and thus yield with the cold starts a tuning-fork shape.

A larger initial entropy acts to shift the final entropy upwards. If the shift is large enough it can push a model that was once in the stalling regime into the cooling regime. An example of this is the case of $\dot{M} = 10^{-3}\ M_\oplus\ {\rm yr^{-1}}$ and $T_0  = 2000\ \mathrm{K}$, which is in the stalling regime for $S_i=9.5\ k_{\rm B}/m_p$ and in the cooling regime for $S_i=11.5\ k_{\rm B}/m_p$.


\section{Summary and Discussion}
\label{sec:summdisc}

In this paper, we investigated the fate of newly accreted matter during the runaway accretion phase of gas giant formation. Since most of the mass of the planet is added during this phase, it is crucial for determining the luminosity of the planet once it reaches its final mass.

\subsection{The Accretion Process}

We showed that solutions for the envelope of an accreting planet take three different forms (\S~\ref{sec:envelope models} and \S~\ref{sec:hot boundary model}) which leads to three different accretion regimes (\S~\ref{sec:acc regimes} and Fig.~\ref{fig:entropy_profiles}). Figure \ref{fig:TP grid scan}
shows the final outcome of accretion: the internal entropy of the planet resulting from accretion with different choices of outer boundary temperature and pressure $T_0$ and $P_0$. The accretion regime depends on the difference between the entropy of the material deposited by the accretion shock $S_0(T_0,P_0)$ and the initial internal entropy $S_i$:

\begin{itemize}
\item The {\em cooling regime}. For $S_0\lesssim S_i$, the planet becomes fully convective, and the superadiabatic gradient drives a large luminosity that leads to rapid cooling. The cooling luminosity is sensitive to the boundary pressure $P_0$, with larger $P_0$ leading to faster cooling. If the cooling is rapid enough compared to the accretion timescale, the end state of this regime is that the internal entropy becomes equal to the surface entropy $S_f\approx S_0$. This regime occurs for low boundary temperatures $T_0\lesssim 500$--$1000\ {\rm K}$.

\item The {\em stalling regime}. For $S_0\gtrsim S_i$, the entropy decreases inwards in a radiative envelope. Provided the entropy contrast is not too great, the envelope joins smoothly onto the interior convection zone. The hot envelope causes the radiative-convective boundary (RCB) to lie at higher pressure than in an isolated cooling planet with the same internal entropy, lowering the luminosity at the RCB and slowing the cooling. In this regime, the final entropy lies close to the initial value of entropy at the onset of accretion $S_f\lesssim S_i$, depending on how much the cooling is slowed. This regime occurs at intermediate temperatures $T_0\approx 1000$--$2000\ {\rm K}$. 

\item The {\em heating regime}. For boundary temperatures $T_0\gtrsim 2000\ {\rm K}$, the entropy difference $\Delta S=S_0-S_i$ cannot be accommodated by the radiative envelope. Instead, the entropy decreases inwards through the envelope to a value $S_{\rm min}>S_i$ (\S~\ref{sec:hot boundary model}, Appendix~\ref{append B}, Fig.~\ref{fig:Smin}) and a second convection zone with entropy $S_{\rm min}$ accumulates on top of the original convective core. Because the minimal entropy $S_{\rm min}$ decreases with increasing planet mass, the envelope quickly moves into the stalling regime as the planet mass increases, and the planet accumulates most of its mass with entropy close to the original $S_{\rm min}$.
\end{itemize}

Our results show that the luminosity of a young gas giant formed by core accretion depends not only on the outer boundary conditions (e.g.~the shock temperature $T_0$) and accretion rate, but also the initial entropy $S_i$ when runaway accretion begins, since it determines whether accretion occurs in the cooling, stalling, or heating regimes. Therefore the thermal state of the young planet in principle provides a link to the structure of the accreting core soon after the crossover mass is reached. This point was also made by \cite{Mordasini2013}, who found that the final entropy depended sensitively on the core mass because it sets the entropy of the envelope at detachment. We see here that for a wide range of intermediate temperatures for which accretion is in the stalling regime ($T_0\approx 1000$--$2000\ {\rm K}$, see Fig.~\ref{fig: ram entropy results}), the final entropy is close to the entropy at the start of runaway accretion.

\begin{figure*}
\epsscale{1.1}
\plotone{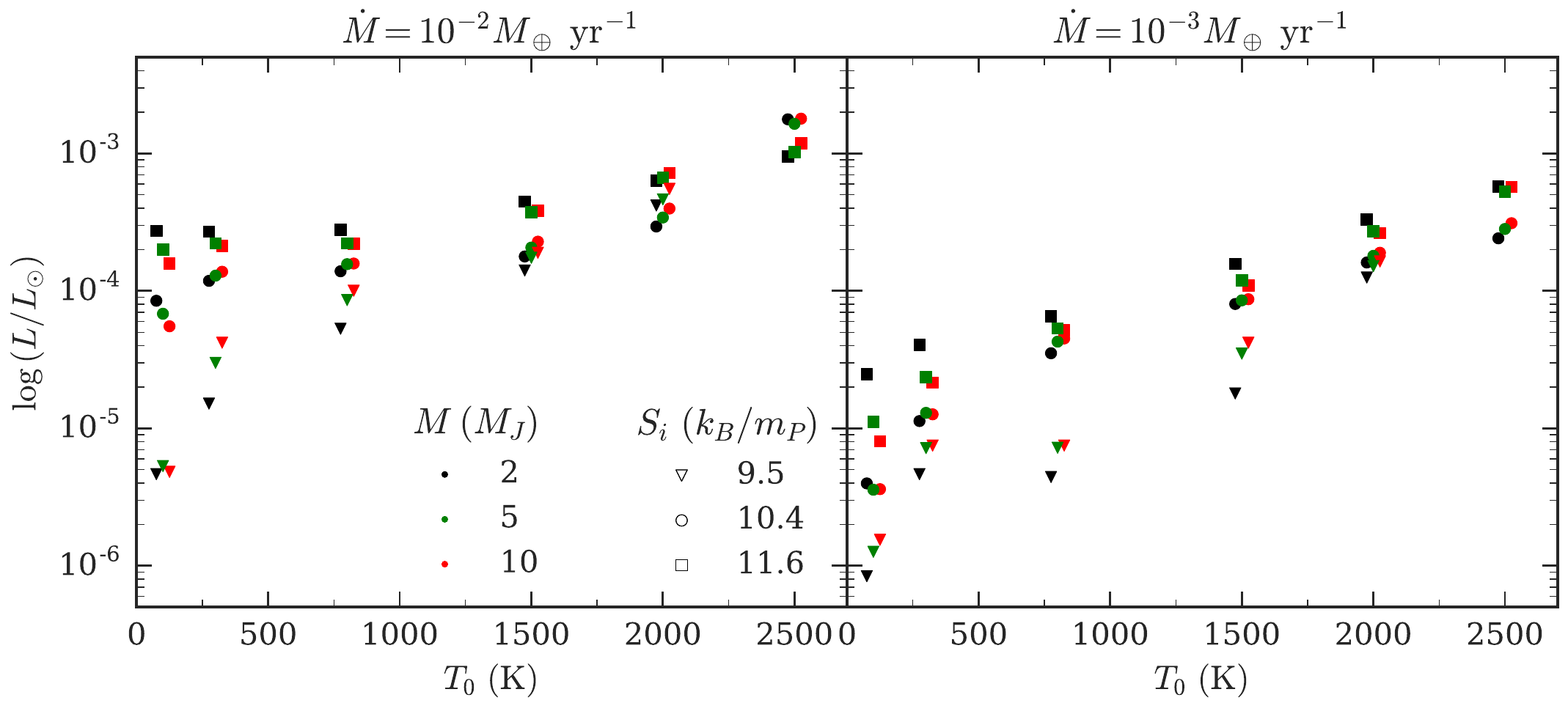}
\caption{Luminosity at the onset of post-accretion cooling as a function of surface temperature during accretion for $\dot M=10^{-2}\ M_\oplus\ {\rm yr^{-1}}$ (\textit{left panel}) or $\dot M=10^{-3}\ M_\oplus\ {\rm yr^{-1}}$ (\textit{right panel}). The colors indicate the final planet mass, while the different symbols indicate the initial entropy of the object at the beginning of accretion (\textit{see legend}). For visual clarity, the markers are given a temperature offset of $-25$, 0, and $+25$~K for a respective final mass of 2, 5, and $10\ M_J$.}
\label{fig: ram pressure cooling curves}
\end{figure*}

\begin{figure*}
\epsscale{1.05}
\plotone{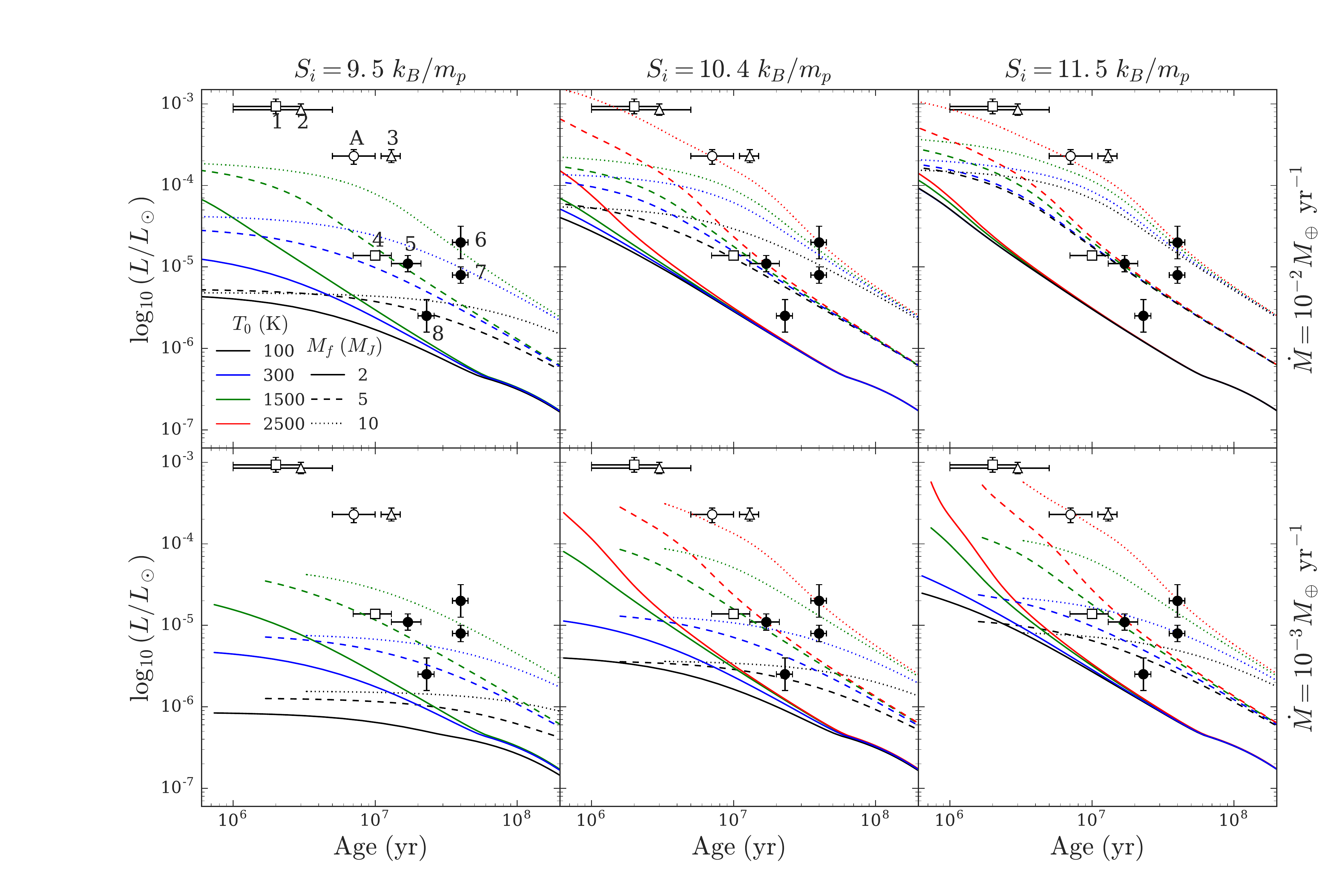}
\caption{Post-accretion cooling compared with directly-imaged exoplanets. The curves show the evolution of the luminosity after accretion ends for final masses $M_f=2$, $5$, and $10\ M_J$ in MESA (\textit{line style}) and surface temperature during accretion $T_0=100$--2500~K (\textit{line color}). \update{The entropy at the beginning of accretion (the accretion rate) is constant along columns (rows); see top (right) titles.} Because these are post-accretion luminosities, the curves begin at different ages based on the total accretion time, which depends on $\dot M$ and the final mass. The data points are for objects with hot-start mass $\lesssim 10\ M_J$ from the compilation of \cite{Bowler2016} as well as the protoplanet HD 100546 b, and use the age \textit{of the host star}: 
\textbf{1}:~ROXs 42B b \citep{Currie2014a}, \textbf{2}:~2M0441+2301B b \citep{Todorov2010}, \textbf{3}:~HD 106906 b \citep{Bailey2014}, \textbf{4}:~2M1207 3932 b \citep{Chauvin2004}, \textbf{5}:~HD 95086 b \citep{Rameau2013}, \textbf{6}:~HR 8799 d \citep{Marois2008}, \textbf{7}:~HR 8799 b \citep{Marois2008}, \textbf{8}:~51 Eri b \citep{Macintosh2015}, \textbf{A}:~HD100546 b \citep{Quanz2015}. 
\update{The symbol type indicates objects around brown dwarfs (\textit{open squares}), objects at $>100\ \rm{au}$ (\textit{open triangles}), planets at $<100\ \rm{au}$ orbiting stars (\textit{closed circles}), and protoplanets (\textit{open circle}).}  
}
\label{fig: cooling curves with data}
\end{figure*}

\subsection{Cold or Hot Starts?}

The luminosity of the planet after formation $L_p$ is shown in Figure \ref{fig: ram pressure cooling curves}. We calculate this luminosity by taking the internal entropy at the end of accretion (for the hot cases, this is the entropy in the hotter, outer convection zone) and constructing a new planet with the same mass and internal entropy in MESA. This avoids convergence issues that arise when changing from accreting to cooling surface boundary conditions at the end of accretion. 

Figure \ref{fig: ram pressure cooling curves} shows that cold starts require that we choose the lowest values of boundary temperature $T_0<300\ {\rm K}$ (comparable to typical nebula temperatures $T_{\rm neb}$), accretion rate $\dot M=10^{-3}\ M_\oplus\ {\rm yr^{-1}}$, and initial entropy $S_i=9.5\ k_{\rm B}/m_p$. In these cases we find luminosities that are comparable to and even lower than the cold-start luminosities of \cite{Marley2007}, who found $2$--$3\times 10^{-6}\ L_\odot$ for $M=4$--$10\ M_J$ and $\approx 6\times 10^{-6}\ L_\odot$ for $M=2\ M_J$. However, increasing any of these parameters beyond these lowest values gives luminosities larger than \cite{Marley2007}. For example, $\dot M=10^{-2}\ M_\oplus\ {\rm yr^{-1}}$ (the limiting accretion rate assumed by \citealt{Marley2007}) gives $L_p\gtrsim 5\times 10^{-6}\ L_\odot$, even for $T_0=100\ {\rm K}$. Increasing $T_0$ beyond $300\ {\rm K}$ gives $L_p\gtrsim 5\times 10^{-6}\ L_\odot$ even for $\dot M=10^{-3}\ M_\oplus\ {\rm yr^{-1}}$.

\update{Temperatures as low as $T_0\sim T_{\rm neb}$ are possible within the boundary prescription of \cite{Bodenheimer2000}, in the case where the flow remains optically thin throughout the growth of the planet. However, the situation in the literature regarding the outer boundary conditions for cold accretion
is somewhat confused. The boundary conditions often used in energy approaches to cold accretion, namely that $L\approx 4\pi R^2 \sigma T_{\rm eff}^4$ and $P_0=(2/3)(g/\kappa)$ (e.g.~\citealt{Hartmann1997,Mordasini2013}, see \S~\ref{sec:previoushotcold}), where $T_{\rm eff}$ is the effective temperature, i.e. the usual boundary conditions for a cooling planet, give temperatures significantly larger than $T_{\rm neb}$. In our models these conditions do not lead to cold starts. The cooling time of the planet is generally longer than the accretion timescale (lower panel of Fig.~\ref{fig:LS}), so that this cooling boundary condition leads to only a small change in entropy during accretion (see the difference between the horizontal solid and dashed lines in Fig.~\ref{fig:center_v_surf}). Only by holding the boundary temperature to a low value are we able to drive a large enough luminosity to accelerate the cooling and reduce the internal entropy significantly on the accretion timescale.}

\update{However, as discussed in \S \ref{sec:previoushotcold}, shock models developed in the context of star formation  \citep{Stahler1980,Commercon2011} and planet accretion \citep{Marleau2016} suggest that the surface temperature is likely to be significantly larger than either of these prescriptions for cold starts. In these models, the gas at the surface of the planet is heated by some fraction of the accretion luminosity generated at the shock to a temperature $T_{\rm{hot}}$ given by $4\pi R^2\sigma T_{\rm{hot}}^4\sim L_{\rm accr} \approx GM\dot{M}/R$. 
In that case our results 
suggest that core accretion will result in hot starts, with high entropy $S_c\sim 12\ k_{\rm B}/m_p$ set by $S_{\rm{min}}$ (\S~ \ref{sec:hot boundary model}) and luminosity $L_p \gtrsim 10^{-4}\ L_\odot$. The planet grows by accumulating hot material on the outside of the original convective core. The entropy $S_{\rm min}$ depends on the accretion rate, but will be difficult to constrain from observed luminosities given the initial rapid cooling for hot starts.}

\subsection{Comparison to Data}

\update{The subsequent cooling of the planets is shown in Figure \ref{fig: cooling curves with data} and compared to measured luminosities of directly-imaged planets. We include those planetary-mass companions listed in Table~1 of \cite{Bowler2016} that are consistent with a hot-start mass $\lesssim 10\ M_J$ (the maximum mass in our models) with ages $\lesssim 10^8\ \rm{yr}$, as well as the protoplanet HD 100546 b which has a bolometric luminosity given by \cite{Quanz2015}. The four points numbered 5--8 refer to planetary companions orbiting at $< 100\ \rm{au}$, and so are perhaps most likely to have formed by core accretion. The cooling curves depend on both $S_i$ and $T_0$ (which set the post-formation entropy), and the planet mass, so that determining the formation conditions is difficult without an independent measurement of the planet mass (e.g.~\citealt{Marleau2014}). Even then, Figure \ref{fig: cooling curves with data} shows that, at the age of these planets ($\approx 20$--40~Myr), the variation in luminosity with shock temperature $T_0$ is less than a factor of a few and can be much smaller for low planet masses and hotter initial conditions.
Younger planets (with ages $\sim 10^6$--$10^7\ \rm{yr}$) have a better memory of their post-formation state. 
However, of the other low-mass objects shown, 
2M~0441~b and 2M~1207~b orbit brown dwarfs, and ROXs~42Bb and HD~106906 are both seen at wide separations (140 and 650~au respectively), so 
it is not clear whether they formed by core accretion.}

\update{The remaining data point is HD 100546 b, which is thought to be a protoplanet that is currently undergoing accretion from the circumstellar disk. The evidence for core accretion, along with its younger age of $\sim 5 \times 10^6\ \rm{yr}$, puts it in the range of planets that will be the most useful in understanding the properties of planets produced by core accretion. Additionally, as previously mentioned in \S \ref{sec: Intro}, it appears that the intrinsic luminosity of the planet can be distinguished from the accretion luminosity, which is an important point to consider when discussing accreting objects. Figures \ref{fig: ram pressure cooling curves} and \ref{fig: cooling curves with data} show that a luminosity of $>10^{-4}\ L_\odot$ is obtained in our models only for hot outer boundaries $T_0\gtrsim 2000\ {\rm K}$ or higher entropies at the onset of runaway accretion $S_i\gtrsim 10\ k_B/m_p$.}

\update{Of all the objects mentioned above,} the need to tune parameters to small values to achieve a cold start has the greatest implications for 51 Eri b, which, with a bolometric luminosity of $1.6$--$4\times 10^{-6}\ L_\odot$ \citep{Macintosh2015}, is perhaps the most likely observed candidate for a cold start. Figure \ref{fig: cooling curves with data} shows that the mass of 51 Eri b could be $10\ M_J$ if $T_0=100\ {\rm K}$, but even a small increase to $T_0=300\ {\rm K}$ requires a lower mass $M\lesssim 3\ M_J$. Therefore it seems likely that the mass of 51 Eri b is close to the hot-start mass, unless the shock temperature can be maintained close to $T_{\rm neb}$ throughout accretion.

\subsection{Future Work}
Our results were obtained holding $T_0$ and $\dot M$ constant during accretion, as the focus of this work was a parameter space study of the effect of particular boundary conditions on the formation of the planet. However, considering a more complex (and realistic) accretion history with time-dependent boundary conditions could result in a different dependence on final mass. For example, the hot-start models produced in our hot accretion regime have a final internal entropy that is relatively independent of planet mass.  This differs from the hot-accretion models of \cite{Mordasini2013}, that show increasing entropy as the planet grows in mass, as in the standard hot-start branch of the tuning-fork diagram (e.g.~compare fig.~2 of \citealt{Mordasini2013} with fig.~2 of \citealt{Marley2007}). \update{Indeed, preliminary work in which we use a surface temperature that depends on the accretion luminosity (as in \citealt{Stahler1980}) shows agreement with traditional tuning fork diagrams for hot starts, i.e.\ an increasing entropy with final mass.}


An additional point related to the consequences of a non-constant surface temperature concerns \S~\ref{sec:acc regimes}, where it was seen that for heating models an outer convective zone made up of the hotter accreted material forms above the initial, lower-entropy core. In the case of constant $T_0$, the planet immediately enters the heating regime, so that at the end of accretion the higher-entropy zone constitutes a large fraction of the mass (95\% in our 10-$M_J$ models).
\update{
However, when $T_0$ is set to the time-dependent $T_{\rm hot}$, it increases with time, and with it the entropy of the accreted material. Therefore, the final internal structure of the planet is different from what is currently seen. This has bearings on the cooling of the object if, for example, an inner radiative region forms \citep{Leconte2012}, but the extent of this effect is presently unclear. A possibility is that thermally irregular internal structures lead to differences even between hot-start cooling curves, implying further uncertainties when estimating the masses of such planets.}  

One of the other goals of this work has been to develop MESA as a tool to study planet formation; we make our \texttt{inlist} and \texttt{run\_star\_extras} files available at \url{http://mesastar.org}. It would be interesting to explore further modelling of gas giant formation in MESA, and overcome some of the limitations of our models. This will require taking into account energy deposition by planetesimals (see review in \S~5.7 of \citealp{Mordasini2015}), modeling the contribution of dust grains to the envelope opacity (e.g.~\citealt{Ormel2014,Mordasini2014b}), including possible composition effects on convection (e.g.~\citealt{Nettelmann2015}), and extending to lower masses than considered here (see \citealt{Chen2016}).

\subsection{Concluding remarks}
We have focused on the runaway accretion phase of gas giant formation and its role in determining the luminosity of young gas giant planets. The results highlight the importance of understanding the physical factors that set the entropy of the planetary embryo while it is still attached to the nebula, and the temperature of the post-shock gas during runaway accretion. \update{This in particular calls for further investigation of the physics occurring directly at the accretion shock, as in \citet{Marleau2016}.} Depending on the shock temperature, the post-formation luminosity spans the full range from cold start to hot start models. This further emphasizes the point made by \cite{Mordasini2013} that large luminosities need not be associated exclusively with formation by gravitational collapse. Beyond the standard core-accretion models, accretion is possibly not spherically symmetric \citep{Lovelace2011,Szulagyi2016,Owen2016}, which also needs to be taken into account.

\update{
We conclude with a few comments pertaining to observations.
Obtaining spectroscopy of young forming objects could significantly help separate the contribution of the shock (also as traced by H\,$\alpha$ as for the LkCa~15 system; \citealp{Sallum2015}) from that of the photosphere. The latter is likely akin to a \mbox{(very-)}low-gravity L/M brown dwarf due to the protoplanet's large radius and surface temperature (see eq.~[\ref{eq:hotT}]).
Also, determining the mass by radial velocity or astrometry, or deriving constraints on it from the morphology of the disk \citep{Bowler2016} would make it possible to break the degeneracy between hot and cold starts \citep{Marleau2014}. Finally, once mass information is available for a sufficient number of directly-imaged planets, it might be feasible to constrain statistically parameters such as the entropy at the beginning of accretion, for instance in the framework of population synthesis \citep{Mordasini2012}.
Thus, exploiting direct-imaging observations by combining them to studies of all factors setting the post-formation thermal state will help constrain the formation mechanism of gas giants.
}


\acknowledgements

The authors would like to thank the referee for comments and insights which helped clarify and improve this paper. DB acknowledges support from a McGill Space Institute (MSI) fellowship as well as a scholarship from the Fonds de Recherche Qu\'eb\'ecois sur la Nature et les Technologies (FQRNT). Additional thanks is given to the participants of the MESA 2016 summer school. AC is supported by an NSERC Discovery grant and is a member of the Centre de Recherche en Astrophysique du Qu\'ebec (CRAQ). GDM was supported in part by a fellowship of the FQRNT and
acknowledges support from the Swiss National Science Foundation under grant BSSGI0\_155816 ``PlanetsInTime''.
Parts of this work have been carried out within the frame of the National Centre for Competence in Research PlanetS supported by the SNSF.

\appendix

\section{The entropy in the envelope}
\label{append A}

\begin{figure}
\epsscale{0.63}
\plotone{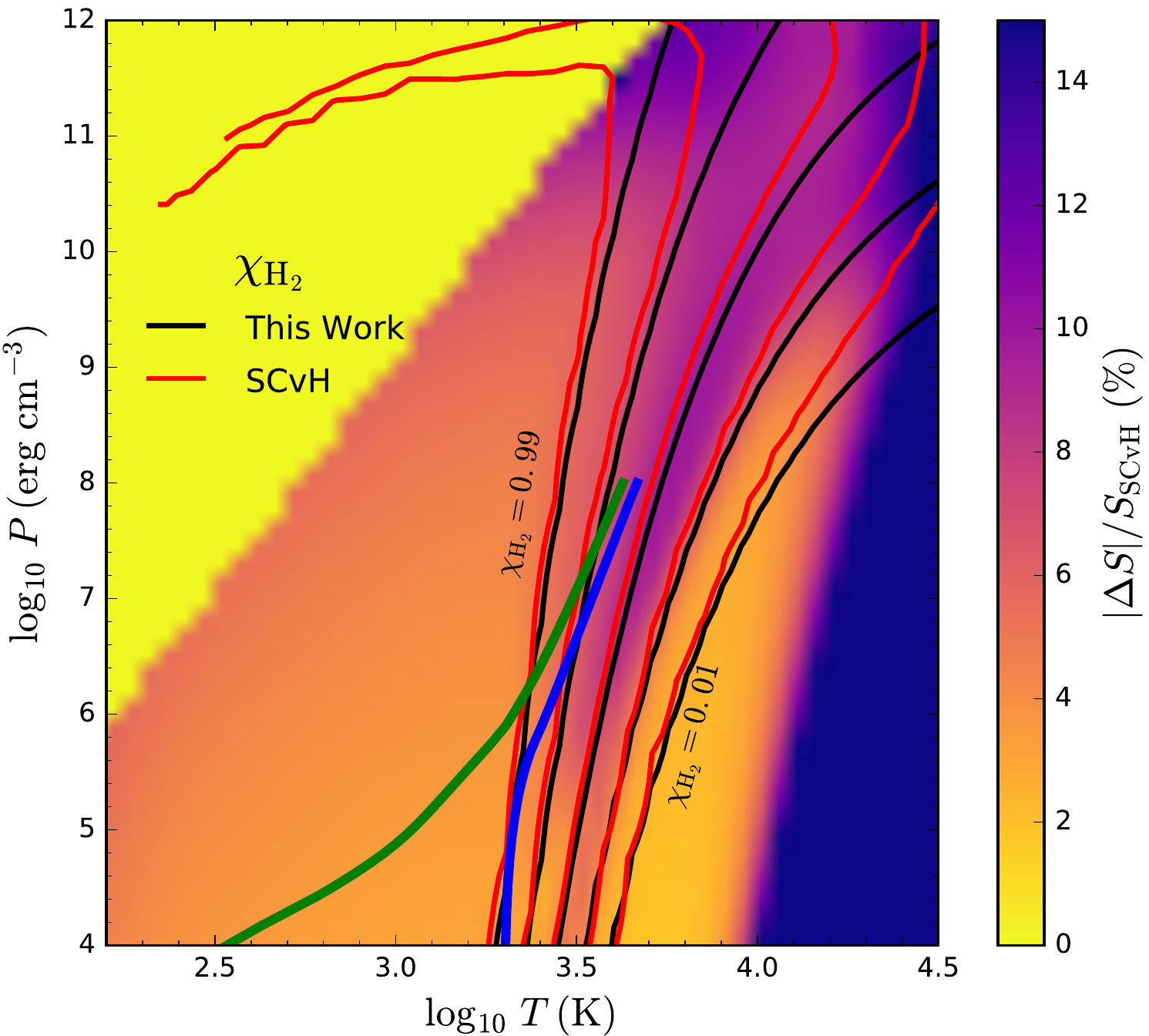}
\caption{Comparison between the entropy calculated using equation (\ref{eq:generic entropy}) and that of \cite{Saumon1995}, $S_\mathrm{SCvH}$. The black and red lines indicate values of constant $\chi_{{\rm H}_2}(P,T)$ obtained using equation (\ref{eq:Saha}) and from \cite{Saumon1995}, respectively, for $\chi_{{\rm H}_2} = 0.01$, 0.1, 0.5, 0.9, 0.99. The thick blue and green lines show envelope models from Figure \ref{fig:profile} with surface temperatures of 2000~K and 150~K respectively. There is no SCvH entropy data in the upper-left, yellow region.}
\label{fig:mesa_vs_calc_entropy}
\end{figure}

In the envelope of the planet, it is a good approximation to assume an ideal gas consisting of molecular and atomic hydrogen as well as helium, in which case we can derive a simple formula for the entropy as a function of pressure and temperature. The ideal gas equation of state is $P=\rho k_{\rm B} T/\mu m_p$ where the mean molecular weight $\mu$ is given by $$\mu^{-1}=\frac{1-Y}{1+\chi_{{\rm H}_2}}+\frac{Y}{4},$$
the molecular fraction $\chi_{{\rm H}_2} = n_{{\rm H}_2}/(n_{{\rm{H}}_2}+n_{\rm H})$ (i.e. $\chi_{{\rm H}_2}$ = 1 (0) is purely molecular (atomic) hydrogen), and $Y$ is the helium mass fraction. The number densities of H and $\rm H_2$ can be computed from the Saha equation
\begin{equation}\label{eq:Saha}
\frac{n_{{\rm H}_2}}{(n_{\rm H})^2} = \frac{n_{Q,{\rm H}_2}z_r}{(n_{Q,{\rm H}})^2}e^{\Delta\epsilon/k_{\rm B}T}
\end{equation}
where $n_{Q,i} = (2\pi \mu_i m_p k_{\rm B} T)^{3/2}/h^3$ and $m_p\mu_i$ is the mass of species $i$. We also consider that for hydrogen gas $n_{{\rm H}_2}+n_{\rm H} = P/k_{\rm B}T$. The ionization energy $\Delta \epsilon$ is $4.48\ {\rm eV}=7.24\times 10^{-12}\ {\rm erg}$ \citep{Blanksby2003} and the rotational partition function for $H_2$ is given by 
\begin{equation}
z_r = \frac{1}{2}\sum_{l=0}^{\infty}(2l+1)e^{-l(l+1)\Theta_{\rm rot}/T},
\end{equation}
which in the limit of $T \gg \Theta_{\rm rot}$ can be approximated as $z_r = T/(2\Theta_{\rm rot})$, where $\Theta_{\rm rot} = 85.4$~K \citep{Hill1986}. The pressure at which a given value of $\chi_{{\rm H}_2}$ is reached at temperature $T$ is
\begin{equation}\label{eq:PfromCHI}
P\left(\chi_{{\rm H}_2},T\right) = 1.6\times 10^6\ {\rm erg\ cm^{-3}}\ \frac{\chi_{{\rm H}_2}}{(1-\chi_{{\rm H}_2})^2}T^{3/2}\exp\left(-{5.4\times 10^4\ {\rm K}\over T}\right).
\end{equation}
Contours of $\chi_{{\rm H}_2}$ in the temperature--pressure plane are shown in Figure \ref{fig:mesa_vs_calc_entropy}. For $T\lesssim 2000\ {\rm K}$ the envelope (pressure range $\approx 10^3$--$10^8\ {\rm erg\ cm^{-3}}$) is molecular, but for higher temperatures atomic hydrogen must be included. 

The entropy per particle of hydrogen and helium is\begin{eqnarray}
  {s_{H_2}\over k_{\rm B}} &=& {7\over 2} + \ln \left({n_{Q,H_2}\over n_{H_2}}\right) +\ln\left({T\over 2\Theta_{\rm rot}}\right)\\
  {s_i\over k_{\rm B}} &=& {5\over 2} + \ln \left({n_{Q,i}\over n_{i}}\right), \quad i = {\rm H}, {\rm He}.
\end{eqnarray}
We use the fact that the temperature is low enough so that the vibrational degrees of freedom of molecular hydrogen, which has a vibrational temperature $\Theta_{\rm vib} = 6210$~K \citep{Hill1986}, are not excited. The entropy per baryon $S_i=s_i/\mu_i$ is then
\begin{eqnarray}
  S_{{\rm H}_2}/k_{\rm B} &=& {1\over 2}\left(20.8 + {5\over 2}\ln T_3 - \ln\rho_{-5}\right),\\
  S_{\rm H}/k_{\rm B} &=& \left(16.3 + {3\over 2}\ln T_3 - \ln\rho_{-5}\right),\\
  S_{\rm He}/k_{\rm B} &=& {1\over 4}\left(19.8 + {3\over 2}\ln T_3 - \ln\rho_{-5}\right),
\end{eqnarray}
where $T_3 \equiv T/1000$~K and $\rho_{-5} \equiv \rho/(10^{-5}\ {\rm g\ cm^{-3}})$. The total entropy per baryon is
\begin{equation}\label{eq:generic entropy}
S/k_{\rm B} = {(1-Y)\over (1+\chi_{{\rm H}_2})}\left[2\chi_{{\rm H}_2}S_{{\rm H}_2}+(1-\chi_{{\rm H}_2})S_{\rm H}\right] + YS_{\rm He} + S_{\rm mix}
\end{equation}
where $S_{\rm mix}$ is the entropy of mixing \citep{Saumon1995} given by
\begin{equation}
  S_{\rm mix}={1\over \mu}\left(-x_{\rm H}\ln x_{\rm H}-x_{{\rm H}_2}\ln x_{{\rm H}_2} -x_{\rm He}\ln x_{\rm He}\right)
\end{equation}
and the number fractions are
\begin{equation}
x_{\rm H} = \frac{(1-Y)(1-\chi_{{\rm H}_2})}{(1+\chi_{{\rm H}_2})}\mu, \quad
x_{{\rm H}_2} = \frac{(1-Y)(2\chi_{{\rm H}_2})}{(1+\chi_{{\rm H}_2})}\frac{\mu}{2}, \quad
x_{{\rm He}} = Y\frac{\mu}{4}.
\end{equation}

Considering the limit of purely molecular hydrogen ($\chi_{{\rm H}_2} = 1$) we find $\mu = 2.28$, $S_{\rm mix} = 0.18$ and the entropy is given by 
\begin{equation}\label{eq:entropy molecular}
{S\over k_{\rm B}/m_p}= 8.80 + 3.38\log_{10} T_3 - 1.01\log_{10}\left({P\over 10^6\ {\rm erg\ cm^{-3}}}\right),
\end{equation}
having used the ideal gas equation of state to rewrite the density in terms of the temperature and pressure. In the other limit of purely atomic hydrogen ($\chi_{{\rm H}_2} = 0$) we find $\mu = 1.23$, $S_{\rm mix} = 0.22$ and
\begin{equation}\label{eq:entropy atomic}
{S\over k_{\rm B}/m_p}= 13.47 + 4.68\log_{10} T_3 - 1.87\log_{10}\left({P\over 10^6\ {\rm erg\ cm^{-3}}}\right).
\end{equation}
From equations (\ref{eq:entropy molecular}) \& (\ref{eq:entropy atomic}) we can read off the adiabatic index $\nabla_{\rm ad} = (\partial \ln T/\partial \ln P)_S = 0.30$ for the molecular case and $\nabla_{\rm ad} = 0.40$ for the atomic case.

In Figure \ref{fig:mesa_vs_calc_entropy} we see how the results of the above equations compare to the values found in \cite{Saumon1995} (SCvH). The blue and green curves, which show envelope models calculated in \S~\ref{sec:envelope models}, are mostly in a region where the deviation from SCvH is only $|\Delta S|/S_{\rm SCvH}\approx 2$--$5 \%$. However, further into the envelope at higher pressures, the error increases to $\sim 10\%$ and so the more detailed equation of state tables from SCvH are required. Large deviations are seen for $T\gtrsim 10^4\ {\rm K}$, where atomic hydrogen is ionized, but this region is not relevant for our envelope models. At lower temperatures, the largest deviations from SCvH occur where $\chi_{{\rm H}_2}$ is transitioning from 0 to 1. Even though our calculation of $\chi_{{\rm H}_2}$ agrees well with that of SCvH (black and red contours in Fig.~\ref{fig:mesa_vs_calc_entropy}), the small differences in $\chi_{{\rm H}_2}$ are amplified in the total entropy because atomic hydrogen gives a much larger contribution to entropy than molecular.

\section{Envelope with power law opacity}
 \label{append B}

We present here analytic solutions for the accreting envelope that clarify why accreting envelopes have a minimum luminosity. We assume a power law opacity of the form $\kappa=\kappa_0 P^\alpha T^\beta$. The equations giving the profile of $T$ and $P$ in the envelope are then
\begin{equation}\label{eq:dTdP}
  {dT\over dP} = {3\kappa_0 P^\alpha L\over 16\pi acGMT^{3-\beta}}
\end{equation}
\begin{equation}\label{eq:dLdP}
  {dL\over dP} = {\dot M c_P T_0\over P}\left(\nabla-\nabla_{\rm ad}\right),
\end{equation}
where in $dL/dP$ we follow \citet{Stahler1988} and assume in the heating term that $T\approx T_0$ is constant in the envelope. Combining equations (\ref{eq:dTdP}) and (\ref{eq:dLdP}) gives a second order ODE for $T$,
\begin{equation}\label{eq:dT2}
  {d^2 T\over dP^2} -{\alpha\over P}{dT\over dP} +{(3-\beta)\over T}\left({dT\over dP}\right)^2 = {3\kappa_0 P^{\alpha}\over 16\pi acGMT^{3-\beta}}{\dot M c_P T_0\over P}\left(\nabla-\nabla_{\rm ad}\right).
\end{equation}
Written in terms of gradients of $\nabla$, this is
\begin{equation}\label{eq:dnabladx}
  {d\nabla\over d\ln P} = (\alpha + 1) \nabla - (4-\beta)\nabla^2 + \gamma\left(\nabla-\nabla_{\rm ad}\right)
\end{equation}
where we define the coefficient $\gamma$ as 
\begin{equation}\label{eq:gamma}
\gamma = {3\kappa_0 P^{1+\alpha}\over 16\pi acGMT^{4-\beta}}\dot M c_P T_0.
\end{equation}

\cite{Arras2006} wrote down equation (\ref{eq:dnabladx}) for non-accreting envelopes, in which case $\dot M=0$ and $\gamma=0$. Starting at low pressure where $\nabla\ll 1$, and assuming $\beta<4$ so that the non-linear term is negative\footnote{For realistic opacities, we find that $\beta>4$ for some regions of the temperature--pressure plane relevant for our envelope models. When $\beta>4$, the non-linear term in eq.~(\ref{eq:dnabladx}) changes sign, and the stable root $\nabla_\infty<0$. A non-accreting envelope with outwards flux $\nabla>0$ at the surface will then always become convective because $\nabla$ increases inwards rapidly. The effect of accretion---to move the isothermal root $\nabla_2$ to a small positive value---is the same whether $\beta$ is smaller or larger than 4. One difference is that a solution with $0<\nabla<\nabla_2$ at low pressure will eventually go to the stable point $\nabla\rightarrow \nabla_1 < 0$ rather than diverging to $\nabla\rightarrow -\infty$. We have also checked the value of $\alpha$ from realistic opacities and find $\alpha>-1$ always so that the linear term in eq.~(\ref{eq:dnabladx}) is positive.}, the solution is that $\nabla$ increases with increasing pressure at first, but eventually saturates at the limiting value 
\begin{equation}
  \nabla_\infty = {1+ \alpha \over 4-\beta}.
\end{equation}
The gradient $\nabla_\infty$ is the radiative zero gradient, for which the first and second terms on the right hand side of eq.~(\ref{eq:dnabladx}) cancel and $d\nabla/dP=0$. \cite{Arras2006} pointed out that the envelope will only become convective at depth if $\nabla$ can exceed $\nabla_{\rm ad}$, i.e. if $\nabla_\infty>\nabla_{\rm ad}$ (so that for example a constant opacity envelope will not become convective since $\nabla_\infty=1/4<\nabla_{\rm ad}$). 

\begin{figure*}
\epsscale{1.0}
\plotone{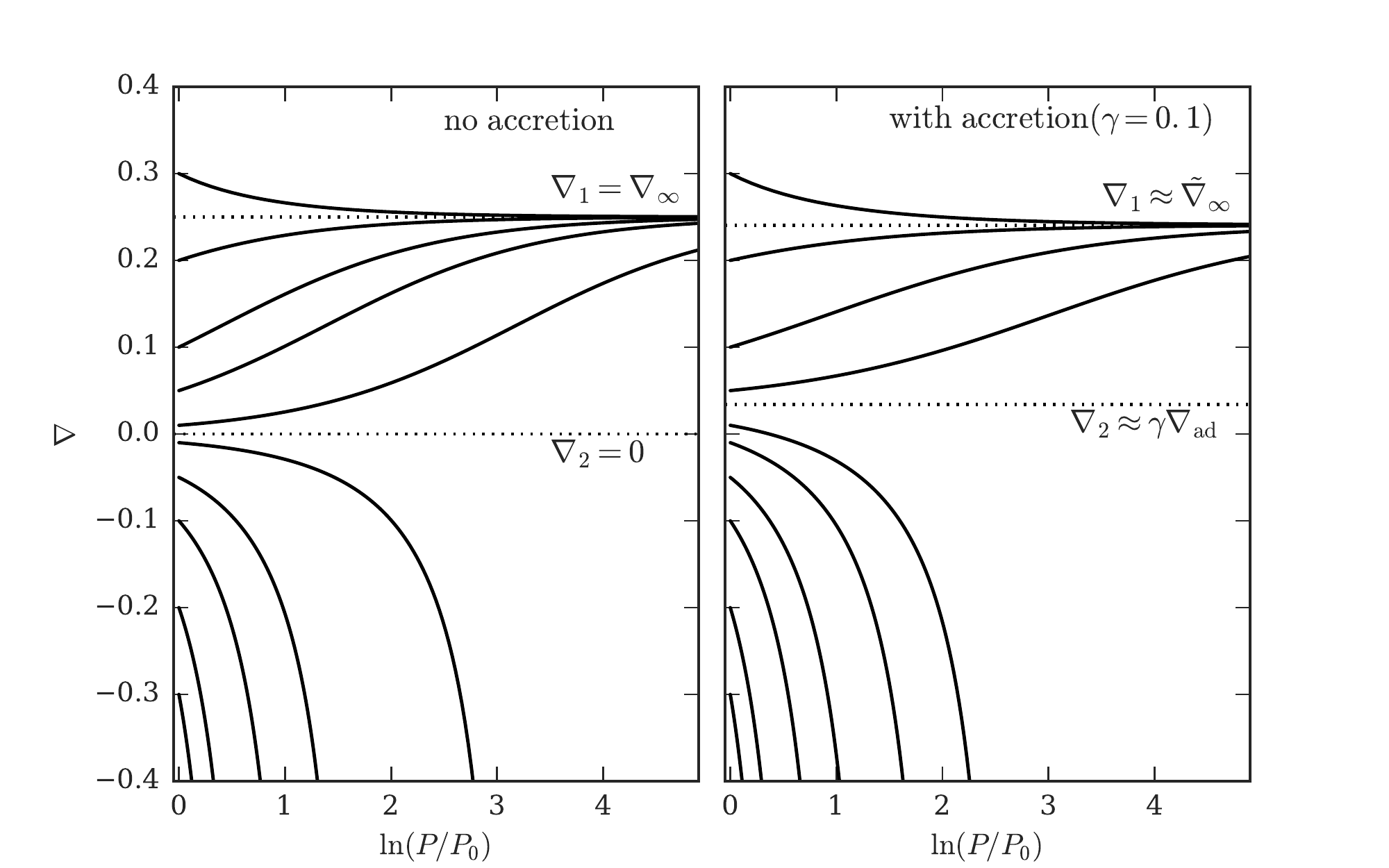}
\caption{Integrations of equation (\ref{eq:dnabladx}) for a constant opacity ($\alpha=0$, $\beta=0$) and $\nabla_{\rm ad}=0.33$, for different initial values $\nabla=\pm 0.01$, $\pm 0.05$, $\pm 0.1$,$\pm 0.2$, and $\pm 0.3$ at $P=P_0$. The dotted horizontal lines show the roots $\nabla_1$ and $\nabla_2$ where $d\nabla/d\ln P =0$. The left panel shows the result with no accretion ($\gamma=0$). The radiative zero slope $\nabla_\infty$ is a stable point; the isothermal slope $\nabla=0$ is unstable. The right panel shows the result with accretion ($\gamma=0.1$). The unstable point now shifts away from zero, becoming $\nabla_2\approx \gamma\nabla_{\rm ad}>0$, corresponding to a minimum luminosity in the envelope.}
\label{fig:toy}
\end{figure*}

Equation (\ref{eq:dnabladx}) has a second root for which $d\nabla/dP=0$, an sothermal envelope with $\nabla=0$. However, an important difference is that, unlike the root $\nabla=\nabla_\infty$, the isothermal solution $\nabla=0$ is unstable. If $\nabla$ is slightly larger than zero, it will increase with pressure and approach the stable solution $\nabla=\nabla_\infty$. If $\nabla$ is slightly less than zero, it will become more and more negative with increasing pressure, $\nabla\rightarrow -\infty$. This is illustrated in the left hand panel of Figure \ref{fig:toy}, which shows the behavior for several different starting values of $\nabla$.

To understand what happens when accretion is included, we consider the case where the coefficient $\gamma$ is a constant (in fact $\gamma$ will be an increasing function of depth for $\nabla<\nabla_\infty$, but we expect this would not qualitatively change the argument). Then setting $d\nabla/d\ln P=0$ gives
\begin{equation}
(4-\beta)\nabla^2 -(1+\alpha+\gamma)\nabla +\gamma\nabla_{\rm ad}=0
\end{equation}
or
\begin{equation}
\nabla = {1+\alpha+\gamma\over 4-\beta}\left[{1\over 2}\pm {1\over 2}\left( 1-{4(4-\beta)\gamma\nabla_{\rm ad}\over (1+\alpha+\gamma)^2}\right)^{1/2}\right]. 
\end{equation}
If we define a modified $\tilde\nabla_\infty=(1+\alpha+\gamma)/(4-\beta)$ this is
\begin{equation}
\nabla = \tilde\nabla_\infty\left[{1\over 2}\pm {1\over 2}\left( 1-{4\gamma\nabla_{\rm ad}\over (1+\alpha+\gamma)\tilde\nabla_\infty}\right)^{1/2}\right].  
\end{equation}
Again we see that there are two roots; for $\gamma\ll 1$ they are
\begin{equation}
\nabla_1 \approx \tilde\nabla_\infty,\hspace{1cm}
\nabla_2\approx {\gamma\nabla_{\rm ad}\over 1+\alpha}.  
\end{equation}
The effect of accretion is to make the unstable root $\nabla_2$ non-zero. An envelope with $\nabla>\nabla_2$ at low pressure will evolve in a stable way: with increasing pressure, $\nabla$ will increase until it reaches the asymptotic value $\nabla=\nabla_1$ (or until it becomes convective and eq.~[\ref{eq:dTdP}] no longer applies). However, if $0<\nabla<\nabla_2$, then $\nabla$ diverges unstably away from $\nabla_2$, decreasing with depth and eventually becoming negative, corresponding to a temperature profile that reaches a maximum and then declines with depth. This is shown in the right hand panel of Figure \ref{fig:toy}, and matches the behaviour we see in our numerical envelope integrations in \S~2 at low luminosity. The luminosity corresponding to the minimum gradient $\nabla_2$ is (from eq.~[\ref{eq:dTdP}] and using eq.~[\ref{eq:gamma}] for $\gamma$)
\begin{equation}
  L_{\rm min} = {16\pi acGMT^4\over 3\kappa P}{\gamma\over 1+\alpha}\nabla_{\rm ad} = \dot M c_P T_0{\nabla_{\rm ad}\over 1+\alpha},
\end{equation}
which is tens of percent of the compressional heating luminosity (cf.\ eq.~[22] of \citealt{Stahler1988} for a similar result).

\end{document}